\documentclass[useAMS,usenatbib]{mn2e}
\usepackage{epsfig,natbib}
\usepackage{amssymb}
\usepackage{graphicx}
\usepackage{placeins}
\usepackage{times}
\usepackage{epsfig}
\usepackage{amsmath}

\def\halpha{\mbox{H$\alpha$}}
\def\hbeta{\mbox{H$\beta$}}

\def\lesssim{\mathrel{\hbox{\rlap{\hbox{\lower4pt\hbox{$\sim$}}}\hbox{$<$}}}}
\def\gtrsim{\mathrel{\hbox{\rlap{\hbox{\lower4pt\hbox{$\sim$}}}\hbox{$>$}}}} 

\newcommand{\kms}{\ensuremath{{\rm km\,s^{-1}}}}
\newcommand\ion[2]{#1{\sc #2}}
\newcommand\fion[2]{$[$#1{\sc #2}$]$}
\newcommand{\hi}{H{\sc i}}
\newcommand{\hii}{H{\sc i} 21cm}

\title[A massive DLA at $z \approx 0.7$]
{ALMA + VLT observations of a Damped Lyman-$\alpha$ absorbing
galaxy: Massive, wide CO emission, gas-rich but with very low SFR
\thanks{Based on observations at ESO; Program IDs:
  095.A-0890(A)(FORS), 079.A-0673(A)(Sinfoni), 274.A-5030(A)(UVES);
  2013.1.01178.S (ALMA). Herschel is an ESA space observatory with
  science instruments provided by European-led Principal Investigator
  consortia and with important participation from NASA.}  }

\author[P. M{\o}ller et al.]{
P. M{\o}ller$^{1}$\thanks{E-mail: pmoller@eso.org}
L. Christensen$^2$,
M. A. Zwaan$^{1}$,
N. Kanekar$^3$,
J.~X. Prochaska$^4$,
N.~H.~P. Rhodin$^2$,
\newauthor 
M. Dessauges-Zavadsky$^5$,
J.~P.~U. Fynbo$^2$,
M. Neeleman$^4$,
T. Zafar$^6$\\
$^1$European Southern Observatory, Karl-Schwarzschildstrasse 2, 85748
Garching bei M\"unchen, Germany\\
$^2$Dark Cosmology Centre, Niels Bohr Institute, Copenhagen
University, Juliane Maries Vej 30, 2100 Copenhagen O, Denmark\\
$^3$ National Centre for Radio Astrophysics, TIFR, Post Bag 3, Ganeshkhind, Pune 411 007, India\\
$^4$ University of California Observatories/Lick Observatory, University of California, Santa Cruz, CA 95064, USA\\
$^5$ Observatoire de Gen\`eve, Universit\'e de Gen\`eve, 51 Ch. des Maillettes, 1290 Sauverny, Switzerland\\
$^6$ Australian Astronomical Observatory, PO Box 915, North Ryde, NSW 1670,
Australia
}

\begin{document}


\pagerange{1 -- 1} \pubyear{2011}

\maketitle


\begin{abstract}
We are undertaking an ALMA survey of molecular gas in galaxies selected
for their strong \ion{H}{i} absorption, so-called DLA/sub-DLA galaxies.
Here we report CO(2-1) detection from a DLA galaxy at $z = 0.716$.  
We also present optical and near-infrared spectra of the galaxy
revealing \fion{O}{ii}, \halpha\ and \fion{N}{ii} emission lines
shifted by $\sim170$ km~s$^{-1}$ relative to the DLA, and providing
an oxygen abundance 3.2 times solar, similar to the absorption
metallicity. We report low
unobscured SFR$\sim1$ M$_{\odot}$~yr$^{-1}$ given the large reservoir of
molecular gas, and also modest obscured
SFR$=4.5_{-2.6}^{+4.4}$M$_{\odot}$~yr$^{-1}$ based
on far-IR and sub-mm data. We determine mass components of the galaxy:
log[${\rm M}_*$/M$_\odot ]= 10.80^{+0.07}_{-0.14}$,
log[M$_{\rm mol-gas}$/M$_\odot ]= 10.37 \pm 0.04$,
and log[M$_{\rm dust}$/M$_\odot ]= 8.45^{+0.10}_{-0.30}$.  
Surprisingly, this HI absorption-selected galaxy has no equivalent
objects in CO surveys of flux-selected samples. The galaxy falls off
current scaling relations for the SFR to molecular gas mass and
CO Tully-Fisher relation. Detailed comparison of kinematical
components of the absorbing, ionized and molecular
gas, combined with their spatial distribution, suggests
that part of the CO gas is both kinematically and spatially
de-coupled from the main galaxy. It is thus possible that a major
star burst in the past could explain the wide CO profile as well as
the low SFR. Support for this also comes from the SED
favouring an instantaneous burst of age $\approx 0.5$ Gyr. 
Our survey will establish whether flux-selected
surveys of molecular gas are missing a key stage in the evolution of
galaxies and their conversion of gas to stars.

\end{abstract}


\section{Introduction}

Since the discovery of the first damped Ly$\alpha$ absorber (DLA) at
high redshift \citep[$z=2.31$; ][]{lowrance72} almost half a century
ago, the nature of the objects causing this absorption has been an
important open problem in galaxy evolution. A wide range of
suggestions, including large galaxy disks, gas-rich dwarf galaxies,
circumgalactic gas, tidally-stripped gas in merger events, and
starburst-driven outflows, have been put forward to explain various
properties of the absorbers
\citep[e.g.][]{prochaska97,haehnelt98,nulsen98,fynbo99,schaye01}.
However, no
single galaxy class has so far been able to explain all the properties
of the known DLA sample. Of course, the absorber sample is likely to
be drawn from the full galaxy population at any redshift and will thus
certainly contain galaxies of different types.

The main limitation in addressing the question of the nature of the DLA
host galaxies has been simply the fact that the bulk of our
information on DLAs stems from absorption studies, which trace only a
pencil beam through the host galaxy. While such studies have provided
detailed information on the \hi\ column density distribution function,
the elemental abundances, the gas-phase metallicities and dust
depletions, the absorption kinematics, the spin temperatures, and the
molecular gas fractions in DLAs at all redshifts
\citep[e.g.][]{prochaska97,pettini97,storrie00,kanekar03,prochaska03,kulkarni05,prochaska05,ledoux06,noterdaeme08,rafelski12,noterdaeme12b,kanekar14},
they have offered limited insight into the link to emission selected
samples. Unfortunately, direct optical imaging of the DLA hosts is
difficult due to their proximity to the bright background quasars,
especially in the case of faint DLA host galaxies. To improve the
chance of detection, one can use the damped profile itself as a natural
coronagraph \citep{smith89,moller93}, or image below the Lyman Limit of
a line of sight neighbour absorber
\citep{steidel92,christensen09,fumagalli15}.

Over the last decade, some progress has been made in understanding the
nature of DLA galaxies via the use of correlations, especially in
absorption properties. For example, DLAs appear to follow
luminosity-metallicity \citep{moller04}, mass-metallicity
\citep{ledoux06,prochaska08} and spin temperature-metallicity
\citep{kanekar09,kanekar14} relations. The use of these correlations
has helped optimize target selection for imaging studies:
specifically, focused observations of high-metallicity DLAs have
resulted in a significant increase in the number of identifications of
$z \geq 2$ DLA hosts in recent years
\citep[e.g.][]{fynbo10,fynbo11,fynbo13,peroux11,peroux12,noterdaeme12,bouche12,krogager13,jorgenson14,hartoog15,zafar17}.
This has allowed a global description of metallicity scaling relations
in DLA host galaxies, and their evolution with redshift
\citep{moller13,neeleman13}, especially the mass-metallicity relation
in DLAs, whose predictions have recently been verified by independent
methods \citep{christensen14}. However, it should be recalled that a
pre-requisite for an imaging study is that the DLA host be bright
enough to be imaged; this effectively limits continuum studies of DLA
hosts, similar to the case of a flux-limited sample. Specifically,
while the HI selection allows to select a sample including low
luminosity galaxies, the
majority of DLA hosts are still undetectable in optical or
near-infrared (IR) imaging studies due to the steep
dependence of luminosity on metallicity \citep{moller04,fynbo08}.

The pre-selection of the highest
metallicity systems for emission searches in many current observing
programmes is thus in reality a flux-limiting sample
pre-selection. While this has significantly increased the efficiency
of those programmes (to a success rate of $\approx50$\% 
\citep{krogager17}), one is
restricted to studies of only a small fraction of the population.
Imaging studies that do not have such pre-selection have a far
lower success rate in imaging the DLA hosts
\citep{charlot91, 
lowenthal95,
lebrun97, 
lanzetta97, 
guillemin97, 
turnshek01, 
warren01, 
moller02, 
colbert02, 
kanekar02, 
lacy03, 
chen03, 
rao03, 
kulkarni06, 
chun06, 
straka10, 
battisti12,fumagalli14,fumagalli15,srianand16}.

Because high metallicity traces more massive, and therefore larger,
galaxies a
side effect of metallicity pre-selection is that the median impact
parameter of the detected hosts is larger for those samples. This
effect was predicted by \citet{moller04,fynbo08} and has been verified
observationally by \citet{krogager12}. Therefore, both long-slit and
integral-field-unit observations of high-metallicity DLA hosts may
miss the target completely if the field of view is small. For example,
\citet{wang15} reported H$\alpha$ non-detections of 3 hosts out to
impact parameters of only $\approx12.5$ kpc and used this to place
limits on their star formation rates (SFRs). However, \citet{krogager17}
investigated this question for several complete samples containing a
large variety of DLA observations, and found that the fraction of
non-detections precisely matches that predicted based on metallicity
and limited field coverage.  Latest, \citet{neeleman17} reported
[\ion{C}{ii}]~158~$\mu$m emission at impact parameters 45 and 18 kpc
from two DLAs at $z \approx 4$, confirming that even at this redshift
the impact parameter will affect detection probability.

Thus, while much work remains to be done in characterizing the
stellar properties of DLA hosts, it is clear that
significant progress has been made in this field over the last decade.
The situation is much worse in studies of the neutral gas that fuels
this star formation, despite the fact that the neutral gas mass of DLA
host galaxies and the gas distribution therein are critical pieces in
understanding galaxy evolution. Unfortunately, the weakness of the
\hii\ line has meant that only a handful of DLAs and sub-DLAs have so
far been searched for \hii\ emission, and all at low redshifts, $z
\lesssim 0.1$
\citep[][]{kanekar01,bowen01,kanekar05,briggs06,mazumdar14,chengalur15}.
So far, only one DLA and one sub-DLA have been detected and mapped in
\hii\ emission \citep{bowen01,chengalur02,briggs06,chengalur15}, at $z
\approx 0.006-0.009$, with fairly tight limits ($\approx \mathrm{few}
\times 10^9 M_\odot$) available on the \hi\ mass of three other systems
\citep{mazumdar14}. Despite this difficulty in measuring the atomic
gas content of individual DLA systems, the connection between DLAs and
\hi\ emission in low redshift galaxies has been made statistically by
\citet{zwaan05}. They show that the characteristics of low-$z$ DLAs
are consistent with their originating in gas discs of galaxies like
those in the $z=0$ population.

Similarly, until recently, the low sensitivity of mm-wave telescopes
has meant that only a handful of searches have been carried out for
molecular gas in emission in DLAs, yielding weak limits on the CO line
luminosity, and thence on the molecular gas mass
\citep[e.g.][]{wiklind94}. The lack of \hii\ and CO emission
detections in DLAs has also meant that we have little information on
the transverse size of the absorbers. The neutral \hi\ column density
can typically only be measured along the sightline to the QSO, except
in the rare cases when a single gravitationally-lensed or double background
QSO can be used to constrain the spatial extent of the neutral gas
\citep[e.g.][]{ellison07,monier09,cooke10,zafar11}.  Mapping of the
\hii\ absorption against the extended radio continuum (and thus
measuring the spatial extent of the neutral gas), has also so far only
been possible in a single DLA, at $z= 0.437$ towards 3C196
\citep{briggs01,kanekar04}.

With the advent of the Atacama Large Millimeter Array (ALMA), it is
now finally possible to directly address the question of the molecular
gas content of DLA host galaxies.  We have hence initiated an ALMA
survey for CO emission in a sample of DLA hosts, to measure their
molecular gas masses, kinematical properties of the gas, determine
the mass fractions in gas and stars,
and their distributions, the star formation efficiencies, and to test
whether DLA galaxies have similar properties as emission-selected
galaxies at similar redshifts.

\begin{table}
\caption{Observing log}
\begin{tabular}{lccccc}
\hline
  &  {\bf ALMA$^1$} \\
\hline
Date  & freq. & bandwidth & channels & resolution & feature   \\
      & (GHz) & (GHz)     & (\#)     & (MHz)      &    \\
\hline
27.12.14 & 132.61 & 2 &  128 & 15.625 & Contin.   \\
 &    134.38 & 1.875 & 3840 &  0.488 & CO(2-1)     \\
 &        144.61 & 2 &  128 & 15.625 & Contin.   \\
 &        146.61 & 2 &  128 & 15.625 & Contin.   \\
\hline
  &  {\bf FORS2} \\
\hline
Date & exp time   & seeing & GRISM & slit & $\lambda / \Delta\lambda$\\
\hline
17.04.15 & 2200 s & 1\farcs2 & 600RI & 1\farcs3 & 900 \\
18.04.15 & 2200 s & 0\farcs9 & 600RI & 1\farcs3 & 900 \\
\hline
\hline
\end{tabular}
[1] The ALMA observations were done with 40 antennas and a maximum
baseline of 348.5 m. On-source time was $\approx$40 minutes, and the
PWV was 5.5 mm.
\label{tab:log}
\end{table}

For our initial pilot ALMA survey, we chose to observe a sample of
high-metallicity DLAs and sub-DLAs, at $z \approx 0.1-0.7$. This
survey has already yielded the first detection of CO emission in a
sub-DLA at $z = 0.101$ towards PKS~0439$-$433 \citep{neeleman16},
originally selected as being part of a QSO-galaxy pair
\citep{petitjean96}. In this paper, we describe our ALMA observations
of a second DLA, at $z_{\rm abs}=0.7163$ towards J1323$-$0021
(SDSS J132323.78-002155.2). The
absorber was originally detected by \citet{rao06} via Hubble Space
Telescope (HST) spectroscopy. A companion galaxy was subsequently
reported as a candidate DLA host by \citet{hewett07}, based on near-IR
imaging data, but attempts to confirm the identification via detection of
optical/NIR emission lines were so far unsuccessful. 

Our ALMA data allow us to definitively identify the DLA host galaxy
via the detection of CO(2-1) line emission, and to characterise
its molecular gas content and distribution.  We also present a new
Very Large Telescope (VLT) FOcal Reducer and low dispersion
Spectrograph (FORS2) spectrum which confirms the identification, as
well as new optimized reductions of archival VLT Spectrograph for INtegral
Field Observations in the Near Infrared (SINFONI) and Ultraviolet
Echelle Spectrograph (UVES) data. In addition re-calibrated Herschel
and HST/STIS data are reduced and analyzed.
The paper is organised as follows: Section~2 describes our new
observations and data analysis, as well as our new reductions of
archival data, Section~3 presents our results and the extracted
physical parameters of the $z = 0.7163$ DLA host galaxy, while
Sections~4 and 5 discuss the new results and summarize our
conclusions. Throughout this paper, we assume a flat $\Lambda$ cold
dark matter ($\Lambda$CDM) cosmology with $H_0=70.4$~\kms~Mpc$^{-1}$
and $\Omega _\Lambda =0.727$ \citep{komatsu11}.

\begin{figure}
\vskip -2.3 cm
\hskip -1 cm
\includegraphics[width=11 cm]{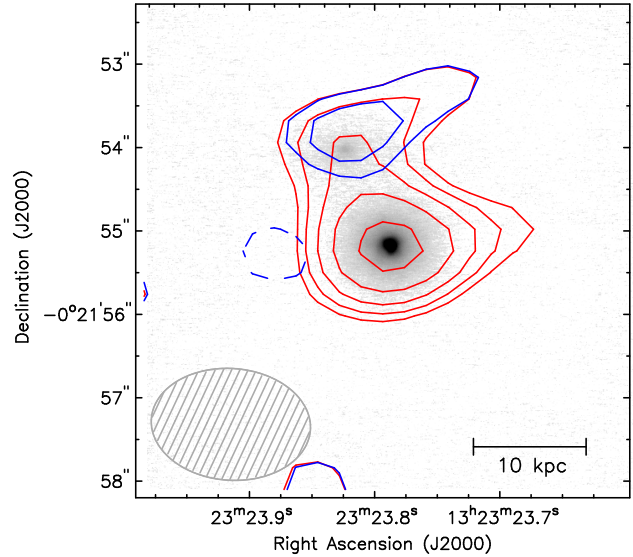}
\vskip -3.3 cm
\caption{
ALMA 140~GHz continuum image (red contours) overlaid on the AO $K$-band
image from \citet{chun10} (kindly provided by Mark R. Chun); the ALMA
beam ($1.91\arcsec \times 1.35\arcsec$) is shown in the lower left corner.
Note the extension of emission to the north of the QSO which matches
the position of the galaxy detected in the infrared image. The contours
start at 2~$\sigma$ and increase by factors of $\sqrt2$; the dashed
contour is at $-2\sigma$. The blue contours are at the same levels, but
are for a continuum image where a point source model at the quasar
position has been subtracted from the U-V visibilities before imaging.
The 140~GHz continuum emission from the galaxy is detected at
$3.2\sigma$ significance.
}
\label{fig:ALMAcontmap}
\end{figure}

\begin{figure}
\includegraphics[width=0.48\textwidth]{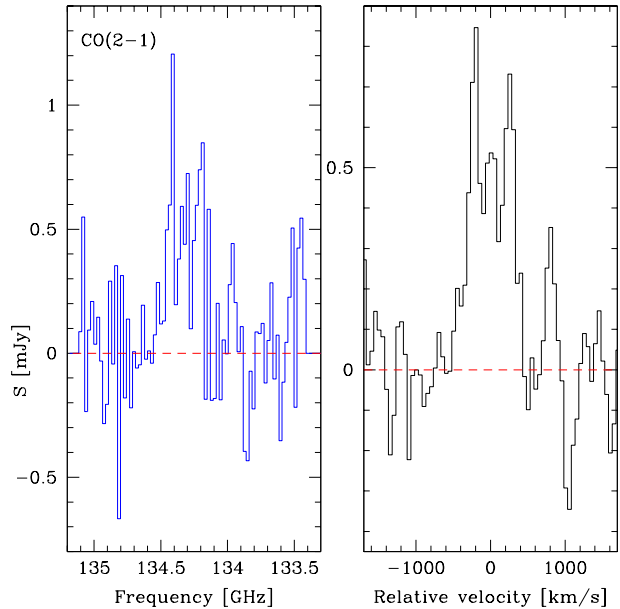}
\caption{The CO(2-1) emission line spectrum: {\bf Left panel:} The
  spectrum at the original velocity resolution of the spectral cube,
  $50$~\kms, with flux density (mJy) plotted against barycentric
  frequency (GHz). {\bf Right panel:} The spectrum slightly smoothed
  (to a resolution FWHM of $\approx 60$~\kms) and here plotted against
  velocity, in \kms, relative to a redshift of $z=0.7163$. 
}
\label{fig:ALMAspectrum}
\end{figure}

\begin{figure*}
\vskip -1.7 cm
\hskip -1 cm
\includegraphics[width=9 cm]{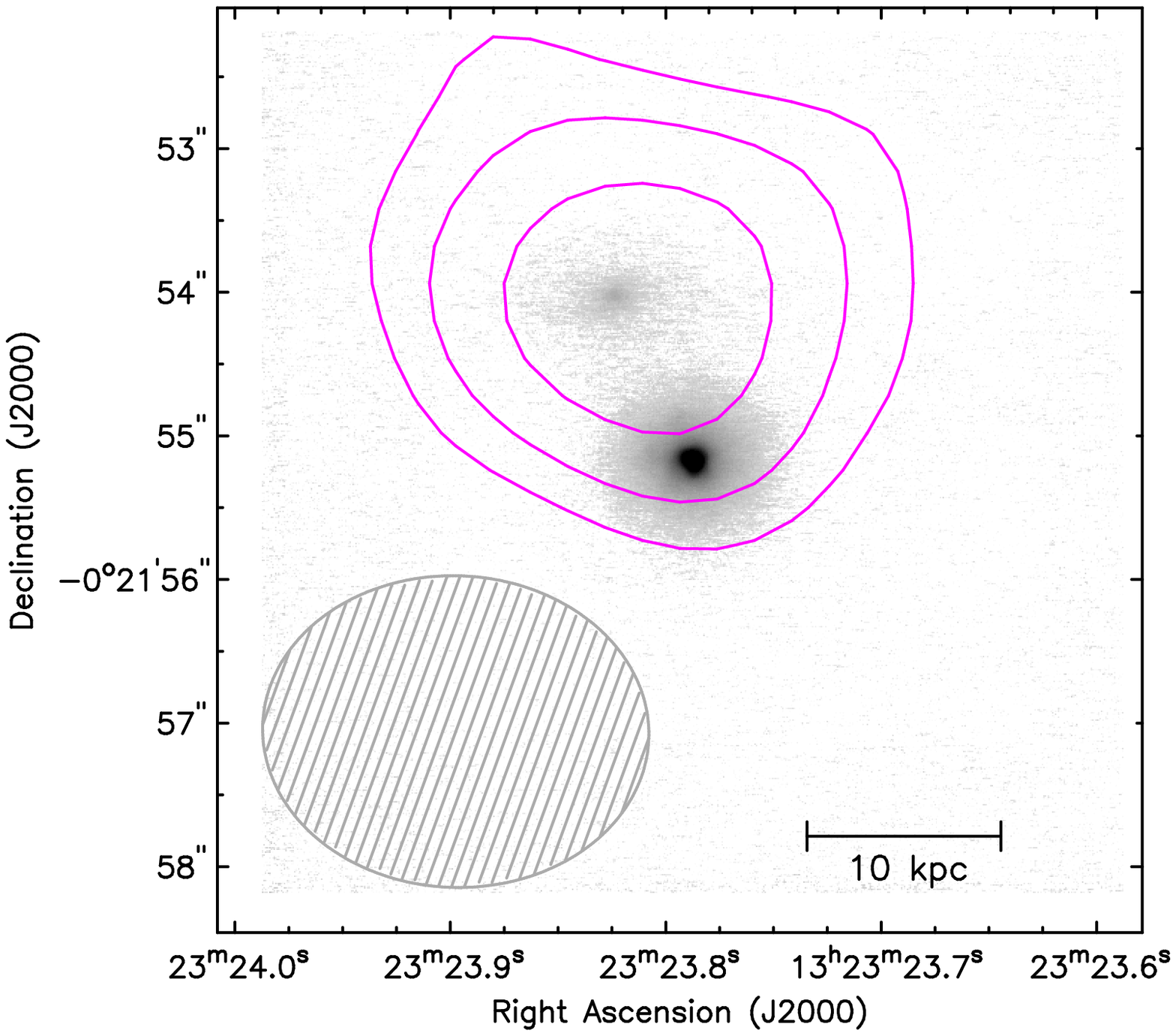}
\includegraphics[width=9 cm]{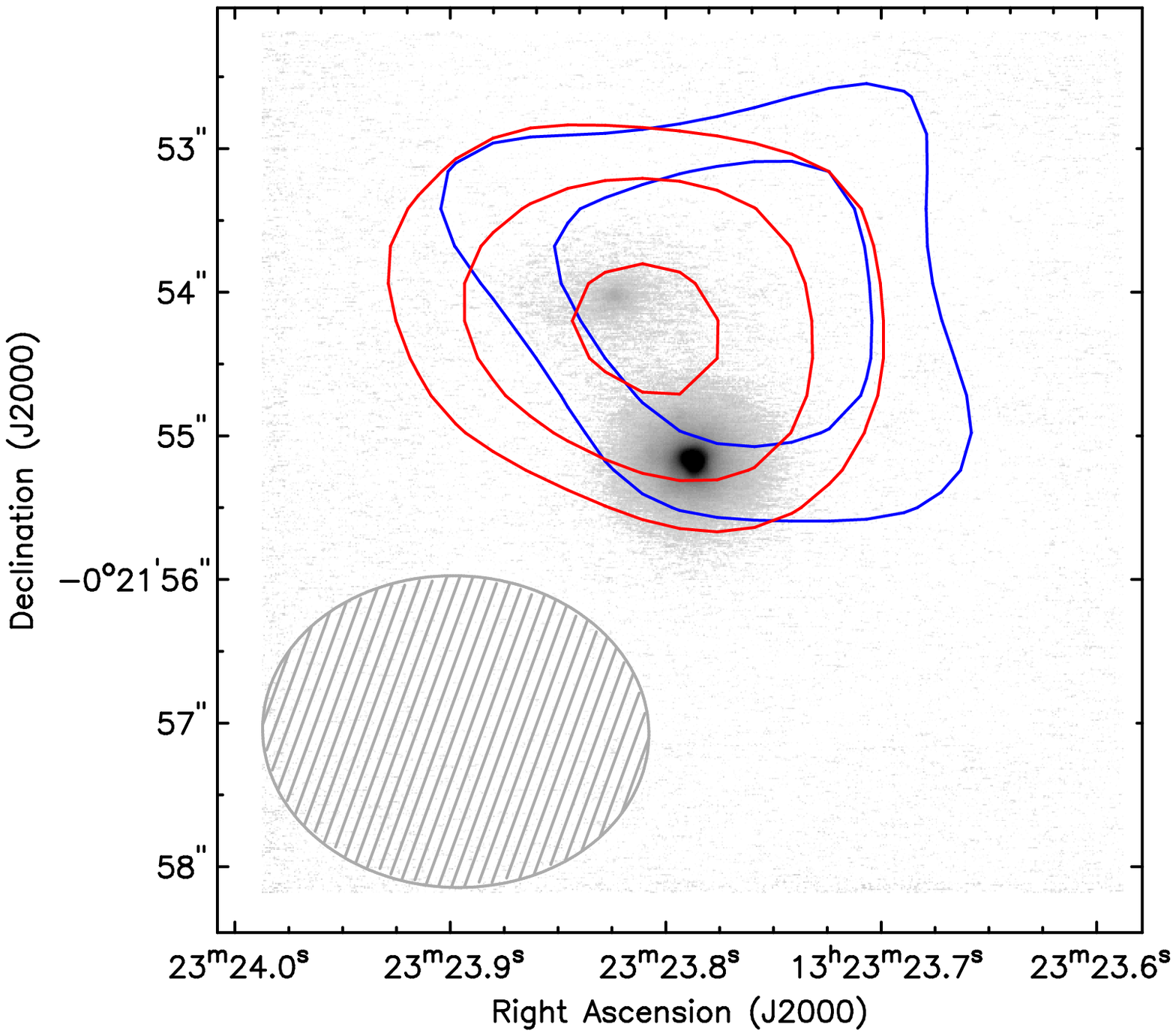}
\vskip -2.6 cm
\caption{{\bf [A]~Left panel:} The total intensity CO(2-1) emission image 
(in purple contours) overlaid on the near-IR $K$-band image of 
Fig.~\ref{fig:ALMAcontmap}. The contours start at 3~$\sigma$ significance 
and increase by factors of $\sqrt2$. The synthesized beam is shown in the 
lower left corner, and is of size $2.7\arcsec \times 2.2\arcsec$.
{\bf [B]~Right panel:} Total intensity CO images of the low
($-400$~\kms\ to $-200$~\kms) 
and high ($-200$~\kms\ to $+400$~\kms) velocity parts of the CO(2-1) 
emission line, in blue and red, respectively, overlaid on the near-IR 
$K$-band image. The contour levels are the same as in the left panel.
It is clear that the high-velocity CO emission is spatially aligned with
the rest-frame optical emission of the DLA host, while the low-velocity
CO emission is slightly offset from the optical emission. Note that the
fact that both maps appear to cover the sightline towards the QSO (the
bright source in the centre of the near-IR image) is likely an artefact
of the synthesized beam.
}
\label{fig:ALMACOoptmap}
\end{figure*}

\section{Observations and archival data}

\subsection{ALMA observations}
\label{sect:ALMA}
The field of the $z= 0.7163$ DLA towards J1323$-$0021 was observed with
ALMA using the band~4 receivers as part of project 2013.1.01178.S, on
27 December 2014 (log provided in Table~\ref{tab:log}). Observing
conditions were below average for
high-frequency studies, with an average precipitable water vapour
measurement of 5.5~mm. The ALMA correlator was configured to use four
spectral windows in dual polarization mode. A single high-resolution
1.875~GHz band, sub-divided into 480 channels, was centred at the
redshifted CO(2-1) line frequency ($134.3$~GHz); this yielded
a velocity resolution of $\approx 16.4$~\kms\ (after online
Hanning-smoothing), and a total velocity coverage of $4185$~\kms\ around
the DLA redshift.
To obtain continuum data the three other bands were used in
low-resolution mode, giving 2~GHz bandwidth and 128 channels each.
Forty 12-metre ALMA antennas were used for the
observations, with a maximum baseline of 348.5~m.  A standard
observing strategy was used, with a phase calibrator (J1319$-$0049)
observed every $\approx 8$~min, and a bandpass calibrator (J1256$-$0547)
observed for 15~min.  Flux calibration was obtained through observing
the quasar 3C279, whose absolute flux scale is frequently verified by
the ALMA observatory. The total on-source time was approximately 40
minutes.

The data editing, calibration and imaging was carried out in the {\sc
  casa} (Common Astronomy Software Applications) package, using
standard procedures.  Stable phase solutions were found for most
antennas, except for a period of $\approx 10$ minutes in the middle of
the observing run. Data for approximately half the antennas were
disregarded for that period. A continuum image of the field was made
by combining data from all four observing bands, excluding the
channels containing line emission. This image was made using
``robust'' weighting, with a Briggs parameter of 0.5, and has a
root-mean-square (RMS) noise level of 16.4~$\mu$Jy/beam.  We clearly
detect continuum emission from the quasar, with a flux density of
$0.14$~mJy.  This is, however, too faint for the purpose of
self-calibration. The continuum map shows a tantalising extension to
the North (see Fig.~\ref{fig:ALMAcontmap}), exactly in the direction
of the candidate galaxy reported 1\farcs1 to the NE of the QSO
\citep{hewett07}, and at 1\farcs25 to the NE in a higher resolution
adaptive-optics image \citep{chun10}. To investigate this further, we
re-imaged the field after subtracting out a point source model at the
quasar position from the U-V visibilities. Both the unsubtracted (red
contours) and the point-source-subtracted (blue contours) data are
shown in Fig.~\ref{fig:ALMAcontmap}. The 140~GHz continuum flux
density at the position of the DLA host galaxy is 53~$\mu$Jy, which
corresponds to a 3.2$\sigma$ detection.  The position of the centroid
of the continuum emission is in excellent agreement with the galaxy
position in the optical image.

A data cube covering the CO(2-1) line redshifted to the redshift
of the DLA was made by imaging the high-resolution visibility
dataset in 50~\kms-wide spectral channels.  Since the data were
obtained with slightly higher spatial resolution than requested, we
tapered the visibilities beyond 75~k$\lambda$ during the imaging, in
order to improve the surface brightness sensitivity. The resulting
spectral cube has a synthesized beam of extent $2.7\arcsec \times
2.2\arcsec$ and an RMS noise of 0.25~mJy/beam per
50~\kms\ channel. The above angular resolution translates to a spatial
resolution of $\approx 19.8 \: {\rm kpc} \times16.1$~kpc at the DLA
redshift, $z = 0.7163$. This relatively coarse spatial resolution
excludes the possibility that we may be resolving out some of the CO
emission.

CO(2-1) emission is clearly detected in the cube, with a peak line
flux density of $(1.70\pm 0.25)$~mJy (i.e. at $\approx 7\sigma$
significance). The CO(2-1) line profile is shown in
Fig.~\ref{fig:ALMAspectrum}; this was obtained by summing all emission
over an aperture of $1.5''$ centred on the optical galaxy. The profile
is seen to display the steep sides and relatively flat central part 
referred to as ``boxy'' profiles or double-horned \citep{davis11},
although the signal-to-noise ratio (S/N) is relatively low.
We obtained a total CO intensity image by adding the flux from all
channels over the central range of $\approx 500$~\kms,
without applying any spatial masking to the channel
maps. Fig.~\ref{fig:ALMACOoptmap}[A] shows an overlay of this total CO
intensity image on the near-IR image; it is clear that the CO emission
is centred on the near-IR emission from the galaxy and not on the
QSO.  With the clear detection of the CO(2-1) emission line, we
identify the neighbouring galaxy as the DLA host: this constitutes 
the first
confirmation of the identification of a DLA host galaxy via molecular
line emission. The velocity-integrated CO(2-1) line flux density is
$S_{\rm int}=(0.50 \pm 0.05)$~Jy~\kms; the measured flux densities are
listed in Table~\ref{tab:summary}.

In addition to the integrated CO image, we also produced two maps where
we separately summed the fluxes from the low velocity CO peak (over the
velocity range $-400$~\kms\ to $-200$~\kms) and the high velocity peak
($+200$~\kms\ to $+400$~\kms). Those are both plotted in
Fig.~\ref{fig:ALMACOoptmap}[B], as blue and red contours respectively.
It is seen that the red set of contours fit the position of the galaxy
well, while the centroid of the blue contours is offset by 0\farcs9
(6.6 kpc) to the West indicating that the molecular gas has a spatially
resolved kinematic structure. We shall return to this in more detail
in Sect.~\ref{sect:kinematics}.

\subsection{Herschel data}
\label{sect:herschel}
The QSO was observed with Herschel/SPIRE \citep{griffin10} on July 31,
2011 in the 250, 350 and 500 $\mu$m bands (OBSID: 1342224987, PI:
V. Kulkarni). We obtained the re-calibrated maps from the Herschel
science archive.

The effective FWHM in the three bands are 17.6, 23.9, and 35.2 arcsec,
respectively, and because of the small separation between the QSO and
DLA galaxy, it was not possible to deconvolve their emission
into two distinct components.  Visual inspection of
the maps showed that the QSO plus DLA galaxy is clearly detected in the
250 and 350~$\mu$m maps, but not in the 500~$\mu$m map, which was
excluded from further analysis.

To measure the integrated flux, the Herschel level~2 maps were divided
by the effective beam area and multiplied by the pixel sizes to get
maps in units of Jy/pixel. We performed aperture photometry following
the description of SPIRE point source photometry in \citet{pearson14}
choosing a small aperture size, and including an aperture correction to
derive the total flux. Errors were obtained using the same apertures on
the Herschel error maps and adding errors in quadrature. The
integrated aperture flux was verified using the data analysis software
'Herschel Interactive Processing Environment (HIPE)' \citep{ott2010}.

The fractional flux detected in the ALMA continuum band from the DLA
galaxy relative to the QSO is $0.27\pm0.07$. To estimate the Herschel
flux from the DLA galaxy alone we assumed that the
fractional flux detected in the ALMA continuum band from the DLA
galaxy is the same also at shorter wavelengths. The
final results are reported in Table~\ref{tab:herschel_phot}.
We discuss, and validate, this assumption in Sect.~\ref{sect:mdust}.

\begin{table}
\begin{tabular}{ll}
\hline
MAP  &  Flux (mJy)\\
\hline
\hline
250$\mu$m  & 11.4$\pm$1.0 \\
350 $\mu$m  & 8.5$\pm$1.1 \\
\hline
\end{tabular}
\caption{Herschel/SPIRE photometry of the DLA galaxy, corrected for
 the fraction of the flux from the QSO (assumed from ALMA data, see
 text). The citet errors include only the photometric errors, not
 possible systematic errors from the QSO fraction correction.
 }
\label{tab:herschel_phot}
\end{table}

\subsection{VLT-FORS2 data}
\label{sect:fors}

\begin{figure}
\begin{center}
\includegraphics[width=0.48\textwidth]{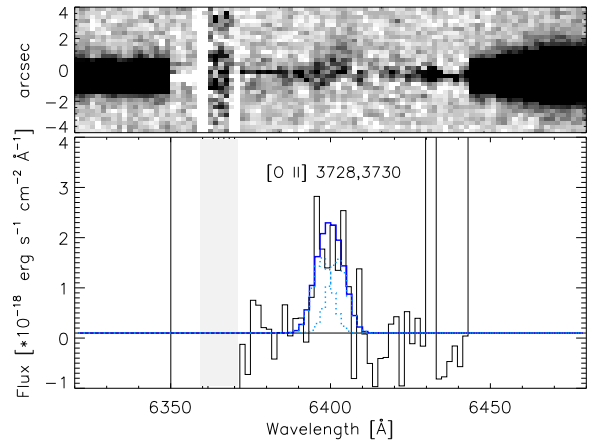}
\end{center}
\caption{{\bf Upper panel:} 
 two-dimensional VLT-FORS2 spectrum around the \fion{O}{ii} emission line
 from the $z=0.7163$ DLA host galaxy. The QSO emission has been
 subtracted in the range $6350-6440$~{\AA}; some residuals
 from this subtraction can be seen in the inner few rows. Excess
 emission is seen at the expected wavelength and impact parameter.
 {\bf Lower panel:} 1D spectrum extracted in the aperture from
 $0.25\arcsec - 2.5\arcsec$ above the QSO trace; The fit of the
 \fion{O}{ii} doublet is overlaid on this spectrum, the two individual
 components as dotted blue lines, the sum as a full blue line.  
}
\label{fig:forsspec}
\end{figure}

The galaxy identified in the ALMA image as the absorber, was
also part of a VLT-FORS2 programme to obtain spectra of DLA galaxy
candidates (programme ID~095.A-0890(A); PI:~L.~Christensen). The galaxy was
observed on 17 and 18 April 2015, with FORS2 long-slit spectroscopy 
using the grism 600RI, which covers the wavelength range $5120 - 8250$~{\AA} 
at a spectral resolution of R$\sim 900$ with the chosen slit width of 
$1.3\arcsec$ . The total integration time was 4$\times1100$ seconds on 
target with the slit aligned at PA$= 42^\circ$ East of North, to cover 
both the QSO and the candidate galaxy based on the relative coordinates 
in \citet{rao11}. The observing conditions were non-photometric, and the 
seeing measured in the spectra varied in time and wavelength over the range
$0.9-1.2\arcsec$, with better values measured on April 18 (Table~\ref{tab:log}).

The VLT-FORS2 data were analysed in the standard manner with bias
subtraction, flat fielding, and wavelength calibration, and sky
background subtraction using a median value from pixels adjacent to
the QSO and the galaxy. To flux calibrate the data, we used an
observation of the spectrophotometric standard star EG274 observed on
17 April. However, as the night was not photometric systematic flux
calibration errors are expected, along with slit losses.  Comparing
the flux-calibrated QSO spectrum with a spectrum obtained in the Sloan
Digital Sky Survey (SDSS) on 19 February 2001, we find that the FORS2
spectrum has a 40\% lower flux level than the SDSS spectrum over the
entire wavelength range. A further comparison of the FORS2 $R$-band
acquisition image with an SDSS $r$-band image reveals that ten bright
stars in the field have the same magnitudes (within 0.01~mag.) in the
FORS2 and SDSS images, while the QSO in the FORS2 image is 0.1~mag
fainter than in the SDSS image. We cannot determine how much the
variation of the flux level in the FORS2 spectra is due to intrinsic
time variations of the QSO flux or slit losses and hence simply report
the measured flux from the DLA host galaxy, with the caveat that the
flux is likely to be low by a factor of $\approx 2$.

A high-resolution Keck $K$-band image, obtained with the use of
adaptive optics, shows a spatial offset of $1\farcs25$ between the
galaxy and the QSO \citep{chun10}. This implies that the two objects
are blended in the moderate seeing of our FORS2 observations. A
careful subtraction of the QSO emission is therefore necessary to
recover any spectral signatures from the galaxy.  To search for the
\fion{O}{ii} $\lambda\lambda$3727,3729 doublet, we assume that the QSO
continuum emission is a pure power law in two wavelength ranges
immediately bluewards and redwards of the expected \fion{O}{ii}
emission feature, and that the spatial point spread function (PSF)
modelled in the adjacent wavelengths does not depend on wavelength.
The measured QSO flux is then scaled to the different spectral
wavelengths and subtracted out, a process known as spectral PSF (or
SPSF) subtraction. The same approach has been used to
find metal emission lines from high-redshift DLA galaxies
\citep{weatherley05,fynbo13,krogager13}. Fig.~\ref{fig:forsspec}
shows the result of the subtraction, which reveals a clear detection
of excess emission at exactly the spatial position of the galaxy.

The \fion{O}{ii} line doublet is not well resolved at the relatively
low FORS2 resolution and we therefore perform a simultaneous fit to
both lines, where we lock the full width at half maximum (FWHM) of
the two lines to be identical, but allow the flux ratio of the lines
to vary freely. We find that the FWHM is $<$300~\kms\ with a heliocentric
wavelength corresponding to \fion{O}{ii} at $z =0.7165 \pm 0.0004$,
where the redshift uncertainty reflects the uncertainty of wavelength
position in the emission line fit from propagated statistical errors.  
There are additional systematic errors from strong subtraction residuals
close to the quasar trace which we take into account as follows. The SPSF
subtraction at the expected galaxy position (from imaging) is excellent
with insignificant residuals, and the extracted \fion{O}{ii} line at
this position 
provides a redshift $z =0.7176 \pm 0.0004$. We take the difference here
to be the result of the sum of systematic and statistical errors, plus
possibly a component from rotation (as suggested from 
Fig.~\ref{fig:forsspec}). Conservatively we assign it all to the
combined error and adopt the average ($z =0.7170 \pm 0.0006$) as the
best redshift estimator where the error spans both extremes.  

The integrated flux from the two components is $(2.8\pm
0.3)\times10^{-17}$~erg~cm$^{-2}$~s$^{-1}$, after correcting for
Galactic extinction of $E_{B-V}=0.024$ \citep{schlafly11}, but not for
intrinsic extinction in the DLA galaxy itself. The full results of the
FORS2 analysis are listed in Table~\ref{tab:summary}.  We note, in
passing, that the low \fion{O}{ii} emission line flux is consistent
with the non-detection of \fion{O}{ii} emission by \citet{straka15}.

\subsection{VLT SINFONI data}
\label{sect:sinfoni}

\citet{peroux11} presented VLT SINFONI $J$-band integral field unit 
spectroscopy of the field of the $z=0.7163$ DLA. They reported a
non-detection of H$\alpha$ emission from the candidate host galaxy.
To complete our multi-wavelength study, we retrieved the SINFONI
dataset (programme ID: 079.A-0673) from the European Southern
Observatory (ESO) archive and re-reduced it. The SINFONI $J$-band
spectra cover the spectral range $11000-14000$~{\AA} at a resolution
of $R \sim 2000$, and the field of view 8~arcsec$\times$8~arcsec. A
total of $13 \times 900$~seconds of integration was split over three
different observing nights in 2007, using a dither pattern for sky
subtraction. A large offset dither pattern avoids nodding into the
galaxy, thus avoiding over-subtraction.

Our data processing was done with the recent version (2.6) of the
SINFONI pipeline; the individual steps involved the same basic
routines used by \citet{peroux11}, apart from the flux calibration and
correction of telluric absorption lines. Specifically, telluric
absorption line corrections may cause differences in our results
compared to the previous reported ones.  We do not initially include a
telluric correction of the data cubes using the calibration
observation of a hot star just before or after the science exposures
as implemented in the pipelines, but instead we derive the correction
directly from the QSO spectrum. We find that this is particularly
important in the spectral region between the $Y$ and $J$ bands, where
the H$\alpha$ emission line from the DLA galaxy falls. Observations on
each of the three observing nights of hot B-type stars with known
magnitudes from the Hipparcos catalogue and known spectral types were
used for flux calibration.

\begin{figure}
\begin{center}
\includegraphics[width=0.45\textwidth]{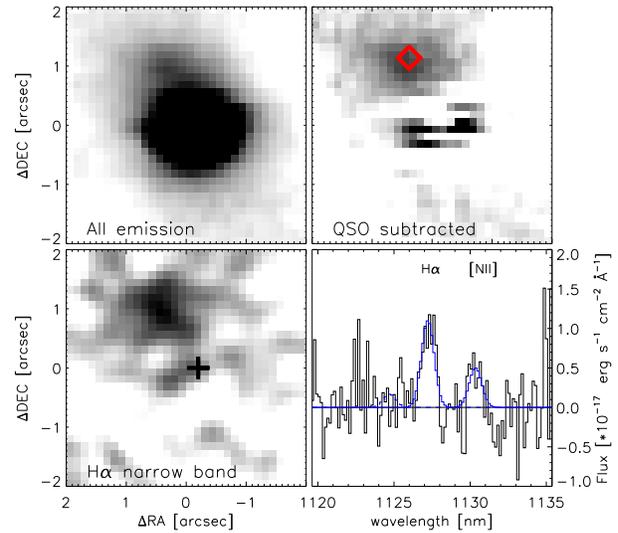}
\end{center}
\vskip -1cm
\caption{Images constructed from the VLT SINFONI data cube as described
 in the text. {\bf Upper left:}
Continuum image with a width of $\approx 50$ nm centred at 13500 nm,
{\bf Upper right:} Same wavelength region after PSF subtraction. The
red diamond marks the centroid of the object seen in the AO $K$-band
image (Fig.~\ref{fig:ALMAcontmap} and \citet{chun10}), {\bf Lower
left panel}: A narrow-band (10\AA ) image centred on the redshifted
\halpha\ line at $z=0.7171$, after subtracting out the continuum
emission; the location of the QSO is marked by a 
  '+' sign. {\bf Lower right panel}: The H$\alpha$+\fion{N}{ii} emission
 line spectrum extracted by selecting spaxels with significant H$\alpha$ 
 emission in the image of the bottom left panel, from which the galaxy
 continuum 
 emission has been subtracted out.}
\label{fig:sinfo_fig}
\end{figure}

To extract information on the galaxy, we first construct an
intermediate-band filter image by summing over an $\approx 50$ nm wide 
region in the data cube centred on 13500~{\AA}.
The upper left hand panel in
Fig.~\ref{fig:sinfo_fig} shows the resulting image, which is clearly
dominated by the QSO emission, but also shows an extension of fainter
emission to the North-East at a distance of
1\farcs2 (equal to the impact parameter of the candidate DLA
host). Fitting a 2-dimensional Moffat function to the QSO location
gives a spatial FWHM of $0.9\arcsec \times 0.7\arcsec$ in the
East-West and North-South directions, respectively.  We then model
the QSO emission by fitting a 2-dimensional Moffat profile with a
spatial location and an FWHM that vary smoothly with wavelength
through the data cube and subtract this model from the initial data
cube. After this, the same spectral region of the data cube is summed
as illustrated in the upper right panel in
Fig.~\ref{fig:sinfo_fig}. Some residuals within the FWHM of the QSO
position are seen, but the continuum emission of the DLA galaxy is now
clearly visible in the image.

We extract a one-dimensional spectrum of the galaxy within an
elliptical aperture of $2\arcsec \times 1.5 \arcsec$ by coadding
spatial pixels (``spaxels'') associated with the galaxy.
After application of corrections for telluric absorption and flux
calibration, the continuum flux level in the chosen region around
$\lambda=13500$ {\AA} is measured to be
$(6.7\pm1.3)\times10^{-18}$~erg~cm$^{-2}$~s$^{-1}$~{\AA}$^{-1}$.
This flux density corresponds to an AB
magnitude of 19.9$\pm$0.25 at this wavelength.
The large uncertainty reflects the added uncertainty from
the QSO subtraction residuals. At the bluest end of the spectral
range, the 1-dimensional spectrum corresponds to a magnitude
$m_{\mathrm{11200}}=20.20\pm0.20$. We include those continuum
magnitudes in our analysis of the
spectral energy distribution of the galaxy in Sect.~\ref{sect:sedfit}.

The extracted spectrum of the galaxy also shows a clear H$\alpha$
emission line at $\lambda=11272.2\pm0.8$~{\AA}. We create a
pure narrow-band H$\alpha$ image by co-adding the relevant wavelength
slices around the emission line and subtracting the adjacent continuum
emission from the galaxy; the result is shown in the lower left panel
of Fig.~\ref{fig:sinfo_fig}. To extract the final, telluric
absorption corrected, emission line spectrum we first define a
$2\arcsec \times 1.5 \arcsec$ rectangular aperture centred on
the DLA galaxy. We sum all spaxels of the original data cube
within this aperture, apply the telluric correction determined
from spaxels only containing QSO continuum, and finally subtract
the continuum from the resulting 1D spectrum. The spectrum is
shown in the lower right hand panel of Fig.~\ref{fig:sinfo_fig}
where we also see a clear
detection of \fion{N}{ii} $\lambda$6584 emission. The other line of
the doublet \fion{N}{ii} $\lambda$6548 is not detected, but this is
consistent with the noise-level of the data. The emission lines
indicate a galaxy redshift of $z=0.7171\pm0.0001$.

Fitting Gaussian functions to the emission lines, we measure a line flux 
of $f(\mathrm{H}\alpha)=(12.0\pm3.9)\times10^{-17}$~erg~cm$^{-2}$~s$^{-1}$ 
and $f$(\fion{N}{ii})=$(5.6\pm1.8)\times10^{-17}$~erg~cm$^{-2}$~s$^{-1}$. 
The H$\alpha$ line is resolved and has an FWHM of $238 \pm 34$~\kms,
after correcting for the instrumental resolution. On both the extracted
J-band ($m_{\mathrm{13500}}$) and narrow-band images we measure
the position of the DLA galaxy relative to the QSO. The resulting
values (PA and impact parameter) are provided in Table~\ref{tab:summary}.

In passing we note that we also detect an emission-line system at
$z=1.38$, at an impact parameter to the QSO sightline of $2.3 \arcsec$ to
the North-East. This system was serendipitously discovered by
\citet{peroux11},
and does not have a corresponding detection in continuum emission
\citep{hewett07,rao11,chun10}.  Additionally, given the high line flux
ratio $\log f($\fion{O}{iii}$)/f(\hbeta)\sim1$ which indicates a hard
ionising source, this system is most likely an extended emission
line region at the redshift of the quasar, and ionized by it. This
emission-line system does not in any way contaminate the results
obtained for the DLA host galaxy.

\subsection{VLT UVES and HST STIS data}
\label{sect:uves}

We obtained the advanced data product VLT-UVES spectrum of the QSO, 
presented in \citet{peroux06}, from the ESO data archive. We corrected 
wavelengths to the heliocentric frame and converted from air to vacuum
wavelengths before combining individual exposures to produce the final
spectrum. The saturated absorption lines in the spectrum, such as
\ion{Fe}{ii}~$\lambda$2344, $\lambda$2382, and $\lambda$2600, show
multiple spectral components at redshifts in the range
$z=0.71531-0.71668$, equivalent to a total velocity spread of $\approx
240$~\kms. Using the definition of $\Delta v_{90}$ \citep{prochaska97},
which is based on the optical depth profile of an unsaturated low
ionization line, we measure a velocity spread of $\Delta v_{90}= 141$~\kms\
from the \ion{Mn}{ii}$\lambda$2576 absorption line
(see Table~\ref{tab:summary}).

The large velocity spread and number of spectral components
makes it difficult to assign a unique redshift to the absorption
system. Also here we then use the definition of $\Delta v_{90}$ 
and assign the median optical depth
($z_{\tau:50}$, the redshift at which 50\% of the optical depth is on
either side) as the effective redshift of the system. Using this
definition we find $z_{\tau:50} = 0.71612$ which implies that the
absorber is kinematically offset from the H$\alpha$ emission by
$171\pm 18$ \kms . In addition we
determine the redshift at which the system has its largest optical
depth, $z_{\tau:max} = 0.71625$.

\citet{peroux06} report that 16 individual components make up the
complex line profile of the DLA and derive a metallicity [Zn/H]~$=
0.61\pm 0.20$, for
log[$N$(\ion{H}{i})/cm$^{-2}$]~$=20.21^{+0.21}_{-0.18}$ (obtained by
\citealt{khare04}).  We note that the identification of the absorber
as a DLA or a sub-DLA and the value of the metallicity is critically
dependent on the \hi\ column density estimate. Specifically,
\citet{rao06} derive log~$N$(\ion{H}{i})~$= 20.54^{+0.16}_{-0.15}$
from the same HST Space Telescope Imaging Spectrograph (STIS)
spectrum.  We therefore downloaded the STIS data from the HST archive
and performed an independent determination of the \ion{H}{i} column
density. Our result agrees with the later analysis of \citet{rao06},
but does not rule out the original result of \citet{khare04}. In the
following analysis, we will use our estimate,
log[$N$(\ion{H}{i})/cm$^{-2}$]~$=~20.4^{+0.3}_{-0.4}$, where the
adopted errors span the full 1$\sigma$ range of the two other
estimates.  This then implies a gas-phase metallicity of [Zn/H]~$=
+0.4\pm 0.3$. We note that \citet{neeleman16HI} did not include this
spectrum in their sample because its S/N is lower than their
threshold of 4.

\section{Results}

\subsection{The molecular gas mass}
\label{sect:molmass}

The velocity-integrated CO J=2-1 line flux density is $S_{\rm
  int}=(0.50 \pm 0.05)$~Jy~\kms.  This can be used to compute the
molecular gas mass of the $z = 0.7163$ galaxy if we know the
CO-to-H$_2$ conversion factor $\alpha_{\rm CO}$ and the nature of the
CO excitation. Both of these depend on the nature of the
galaxy: for example, spiral disk galaxies like the Milky Way have
$\alpha_{\rm CO} \approx 4 \: {\rm M}_\odot$~(K~\kms~pc$^{2}$)$^{-1}$,
while QSOs and starburst galaxies (e.g. ultraluminous infrared
galaxies and sub-millimetre galaxies) have far lower values,
$\alpha_{\rm CO} \approx 1 \: {\rm M}_\odot$~(K~\kms~pc$^{2}$)$^{-1}$,
both after including a factor of 1.36 to correct for the contribution
of helium \citep[e.g.][]{bolatto13,carilli13}. The conversion factor
also has a strong dependence on metallicity, with far higher
$\alpha_{\rm CO}$ values
($\gtrsim 30 \: {\rm M}_\odot$~(K~\kms~pc$^2$)$^{-1}$)
obtained in low-metallicity
systems \citep[e.g.][]{leroy11} than in high-metallicity galaxies
($\alpha_{\rm CO} \lesssim 4 \: {\rm
M}_\odot$~(K~\kms~pc$^2$)$^{-1}$).
In the present case, the high
metallicity of the DLA host galaxy indicates a low $\alpha_{\rm CO}$
value.

Next, CO in QSOs and starburst galaxies has typically been found to be
thermalized up to the J$=3$ level, with sub-thermal excitation of the
higher-J levels \citep[e.g.][]{weiss07,danielson11}. Conversely, even
the J$=2$ level is typically sub-thermally excited in spiral galaxies
\citep[e.g.][]{fixsen99,dannerbauer09,carilli13}.  In thermal
equilibrium, the brightness temperature of the CO J$=2-1$ line would
be equal to
that of the J$=1-0$ line, while in the Milky Way, the J$=2-1$
brightness
temperature is $\approx 0.6$ times that of the J$=1-0$
line. We will assume sub-thermal
excitation of the J$=2$ level and a brightness temperature line ratio
of $r_{21}
\approx 0.6$ ($r_{21}=L'_{\rm CO(2-1)} / L'_{\rm CO(1-0)} =
S_{\rm CO(2-1)} / (4\times S_{\rm CO(1-0)})$)
 for the J$=2-1$ and J$=1-0$ lines, consistent with the
assumption of a Milky-Way value for $\alpha_{\rm CO}$.

For the assumed $\Lambda$CDM cosmology, the CO J$=2-1$ flux density of
$S_{\rm int}=(0.50 \pm 0.05)$~Jy~\kms\ implies a CO J$=1-0$ flux
density of $(0.20 \pm 0.02)$~Jy~\kms, and thence, a CO~J$=1-0$ line
luminosity of $L'(1-0) = (5.55 \pm 0.55) \times
10^9$~K~\kms~pc$^2$. Using $\alpha_{\rm CO} = 4.2 \: {\rm
  M}_\odot$~(K~\kms~pc$^2$)$^{-1}$
\citep[recommended by][]{bolatto13} 
then yields a total molecular gas mass of
M$_{\rm mol-gas} = (2.33 \pm 0.23) \times [\alpha_{\rm CO}/4.2]
\times (0.6/r_{21}) \times 10^{10} \: {\rm M}_\odot$ for the DLA host
galaxy.  Conversely, if we were to assume that the DLA host contains
an obscured starburst,
with $\alpha_{\rm CO} \approx 1$ and $r_{21} \approx 1$ (i.e. that the
J$=2$ level is thermalized), we obtain a molecular gas mass of
M$_{\rm mol-gas} = (3.5 \pm 0.35) \times [\alpha_{\rm CO}/1.0] \times
(1.0/r_{21}) \times 10^{9} \: {\rm M}_\odot$.  
Since the SFRs estimated from both the H$\alpha$ line and the dust
continuum are low ($\lesssim 5$~M$_\odot$~yr$^{-1}$ see Sections 
\ref{sect:sfr} and \ref{sect:mdust}), we consider the starburst
scenario to be unlikely. This is further supported by the
galaxy's gas-to-dust ratio (see Sect.~\ref{sec:gastodust}). We therefore
conclude that the most likely scenario is the former, and that the 
molecular gas mass of the DLA host galaxy is likely to be very high.  
The values of $\alpha_{\rm CO}$ applied here already include
the correction (1.36) for He.

\subsection{Kinematics of the molecular, neutral and ionized gas}
\label{sect:kinematics}

\begin{figure}
\begin{center}
\hskip -6 mm
\includegraphics[width=0.51\textwidth]{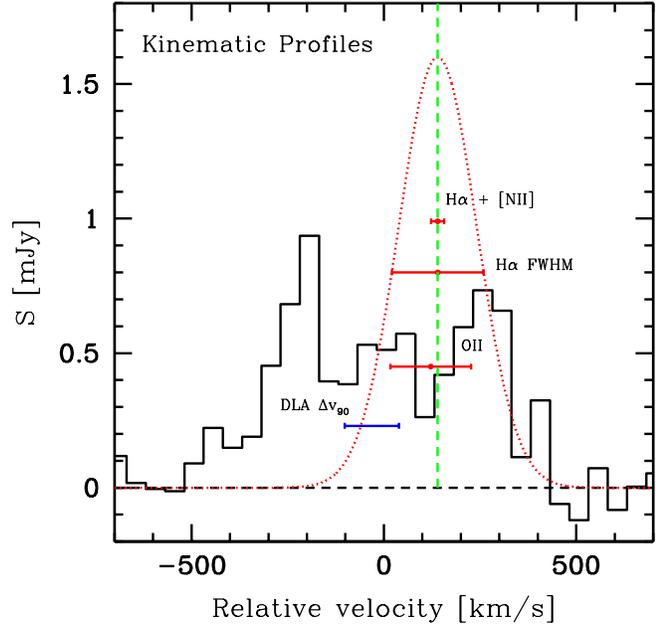}
\end{center}
\vskip -0.5 cm
\caption{Detailed kinematical structure of the three gas components:
molecular (black histogram), neutral (in absorption, blue horizontal bar),
and ionized (red). All velocities are relative to $z=0.7163$.
The rest-frame optical emission lines are seen to be aligned with the
central and the highest redshift peak of the molecular gas.
Error bars on H$\alpha$ + \fion{N}{ii} and
\fion{O}{ii} points are measurement errors while the bar marked
``H$\alpha$ FWHM'' spans the FWHM of the H$\alpha$ line.  
To assist the comparison
we overplot the Gaussian fit to the H$\alpha$ line from
Fig.~\ref{fig:sinfo_fig}.
}
\label{fig:dynamics}
\end{figure}

In Fig.~\ref{fig:dynamics}, we provide an enlarged view of the CO
J$=2-1$ emission line profile shown in the right panel of
Fig.~\ref{fig:ALMAspectrum}. The flat-topped appearance of a
``boxy'' profile \citep{davis11} is here evident. The flat profile 
appears to consist of three individual CO velocity components, but
the details of this sub-structure is
difficult to verify at the current S/N and must await better data.
Overlaid in Fig.~\ref{fig:dynamics} we show the spectral components of
the ionized gas seen in the VLT-SINFONI and VLT-FORS2 emission spectra
(marked in red), and the absorbing DLA gas seen in the VLT-UVES spectrum
(blue). For the absorption profiles, we mark the velocity
spread containing 90\% of the integrated optical depth of
low-ionization unsaturated metal lines, $\Delta v_{90}$.

Curiously, Fig.~\ref{fig:dynamics} indicates that the H$\alpha$ emission
is kinematically related only to the two highest redshift molecular
components, while the lowest-redshift molecular feature appears to be
kinematically distinct from both the optical emission and absorption
lines. The lowest-velocity molecular peak spans the bins from which
we produced the blue contour map in Fig.~\ref{fig:ALMACOoptmap}, where it
was seen that this peak also appears spatially offset from the optical
galaxy. The velocity span of the metal-line absorption lies
kinematically between the reddest and bluest CO velocity components,
which are plotted as red and blue contours in Fig.~\ref{fig:ALMACOoptmap}.

\subsubsection{Centroid positions of emission sub-components}
\label{sect:centroid}

In order to address this issue of position of the various emission
sub-components quantitatively, we determine the positions of the dust
and of the red and blue CO components. For DLA hosts, this information
is traditionally provided in the form of the position angle (PA) and 
the impact parameter ($b_{\rm DLA}$), relative to the quasar. Based
on the high resolution $K$-band data, \citet{chun10} reported
(PA, $b_{\rm DLA}) = (24.5^\circ, 1\farcs25)$. From the images of the
dust continuum, the red CO component, and the blue CO component, we
obtain (PA, $b_{\rm DLA}$)~$= (17\pm8^\circ, 1\farcs46\pm0\farcs2)$,
$(32\pm14^\circ, 1\farcs03\pm0\farcs3)$,
and $(-18\pm14^\circ, 1\farcs15\pm0\farcs3)$, 
respectively. It is apparent that the first two sets of values agree
well with the optical position of \citet{chun10}, whereas the last
does not. More specifically, measuring the distance between the
$K$-band position of the host and the centroid of the image of the
blue CO component, we find $0\farcs9\pm0\farcs3$. This is statistically
significant evidence that the blue CO component is spatially
offset from the other three (optical, dust, red CO) components.

From the results summary in Table~\ref{tab:summary} we see that
there is agreement (to within the errors) between the relative
position of the DLA galaxy detected in $K$-band, $J$-band, dust
continuum, H$\alpha$ and ``red CO component'', i.e. all of those
emission components of the Damped absorber appear to be centred
on the optical galaxy.

\subsection{The CO line width and the dynamical mass}
\label{sect:W20W50}

\begin{figure}
\begin{center}
\includegraphics[width=8.4 cm]{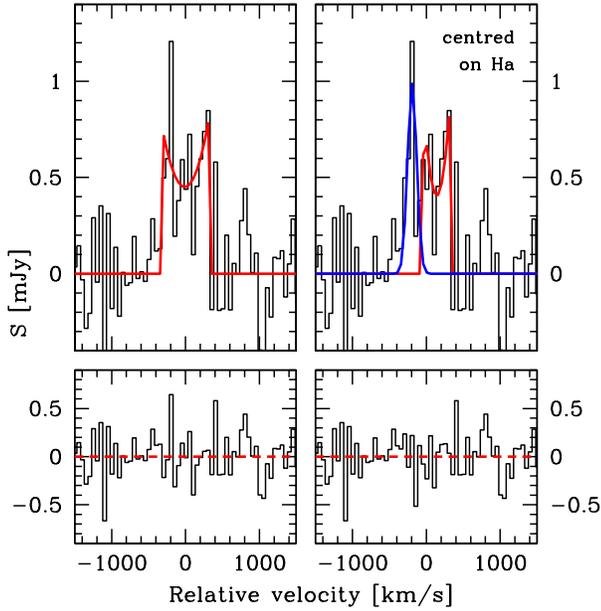}
\end{center}
\caption{{\bf upper panel, left:} Double-horn profile fit to the CO(2-1)
line (red); {\bf upper right:} fit using the same double-horn profile
function but forced to be centred at the same redshift as the H$\alpha$
line (red), plus a single gauss profile fit to the remaining peak (blue).
{\bf Lower panels:} residuals.} 
\label{fig:2hornfit}
\end{figure}

As noted in Sect.~\ref{sect:ALMA} the CO J=2-1 line profile matches
the definition of a boxy profile. To obtain measures for the line width
which are directly comparable to those of the COLD GASS galaxy sample
\citep{saintonge11} we use the same fitting formula for a double-horn
profile \citep[equation A2 in][]{tiley2016}. A fit of this function
to the unsmoothed data (shown
in Fig.~\ref{fig:2hornfit}, left panel) finds a CO central redshift of
$z = 0.71645(9)$, a W50 of $647\pm54$
\kms\ and W20 of $652\pm54$ \kms\ (uncorrected for inclination).  
Our CO map does not have the resolution to obtain an inclination angle of
the disk, but from the high resolution AO image presented by
\citet{chun10} (their figures 1 and 2) we estimate that the inclination
is in the range 40-48$^o$ (including also the effect of a slightly
elongated residual PSF). Using the most conservative value we get
W50/sin(48$^o$) = 871 \kms .

This is an extremely large velocity width for CO emission, comparable
to those obtained in the most massive sub-millimetre galaxies and
QSOs \citep[e.g.][]{carilli13}.
As noted above, it is possible that the bluest feature of the
profile may be due to a separate object which is kinematically detached
from the main galaxy. In particular the H$\alpha$ line is offset from
the CO central redshift. To test this possibility we also perform a
second fit using the same functional form, but now locking the central
frequency of the fit to that of the H$\alpha$ line. The fit is
restricted to the spectral range from $-200$ \kms\ and above, and
gives  W50 = $378\pm52$ \kms , W20 = $384\pm52$ \kms\ which is more in
line with typical galaxies of similar stellar mass. In addition we fit a
single gauss profile to the remaining blue peak which covers the spectral
range used for the blue contours in Fig.~\ref{fig:ALMACOoptmap}.
The best fit gauss has $\sigma = 61$ \kms\ and an offset of $-212$ \kms\
relative to the fiducial DLA redshift (0.7163), resulting in a total
offset between this blue component and the H$\alpha$ line of 352 \kms .
The combined fit is shown in the right panel of Fig.~\ref{fig:2hornfit}.
The flaring parameter ($R_{2,5}=$W20/W50) is found to be 1.01 and
1.02 for the two fits respectively, confirming the profile shape as
boxy ($R_{2,5}<1.2$).

The dynamical mass of the DLA host can be roughly estimated as M$_{\rm dyn}
\sim (\Delta V^2 \times R)/G$, where $\Delta V$ is the characteristic 
velocity, and $R$ the characteristic size. We will assume that the entire 
CO emission arises from a single object. A lower limit on its spatial 
extent can be obtained from the transverse separation 
between the optical and the blue CO components; the angular 
separation of $0\farcs9$ corresponds to a spatial separation of 
$\approx 6.5$~kpc.
Assuming $\Delta V \approx ({\rm W50}/{\rm sin}(48^o ))/2 =
436$~\kms\
and $R \gtrsim 6.5$~kpc, we obtain M$_{\rm dyn} \gtrsim 2.9 \times
10^{11} \: {\rm M}_\odot$.

\subsection{Results from the optical and near-IR data}

The excellent spatial alignment between the CO emission map and the
optical image of the candidate host galaxy in
Fig.~\ref{fig:ALMACOoptmap}A, 
and the good agreement between the CO and the metal-line absorption and 
emission velocities in Sect.~\ref{sect:kinematics} clearly demonstrate
that the CO J$=2-1$ emission originates in gas associated with
the galaxy detected in the
$K$-band and optical images.  This constitutes the first case of
spectroscopic identification of a DLA host galaxy via its molecular line
emission.  This is an important result in its own right, because
it shows that ALMA has opened up a new window for the identification 
of DLA hosts, a task that has been fraught with much difficulty in 
the past. Of course, once the host has been identified via its molecular 
emission lines, additional data can be obtained and already existing
data can be exploited, as has been done here.

\subsubsection{The stellar mass of the $z = 0.7163$ galaxy}
\label{sect:sedfit}

To derive the stellar mass of the $z=0.7163$ galaxy, we use the
standard approach of fitting to the spectral energy distribution
(SED), using release version 12.2 of {\sc Hyperz}
\citep{bolzonella00}. We assume a Chabrier initial mass function
(IMF), and derive different stellar masses, using templates with
various star formation histories and metallicities between 0.2 and
2.5 times solar values \citep[see][for further details]{christensen14}.

The input photometry of the galaxy in $ugri$ filters is taken from
\citet{rao11}.  However, strong residuals from the subtraction of the
QSO's point spread function have a strong effect on the $i$-band
measurement \citep{rao11}\footnote{Images can be found at
  http://enki.phyast.pitt.edu/Imaging.php} and we hence choose to not
include this band in the SED fit. The near-IR $K$-band data are from
\citet{hewett07}, who obtained a magnitude similar to that obtained
from the $K$-band adaptive optics data \citep{chun10}.  In addition,
we also use the magnitudes measured from our VLT-SINFONI data, derived
in section~\ref{sect:sinfoni}. All photometric data are listed in
Table~\ref{tab:phot}.

\begin{table}
\begin{tabular}{lll}
\hline
\hline
Filter & Magnitude  & Reference \\
\hline
$u$  &  24.89$\pm$0.81  & [1] \\
$g$  &  22.64$\pm$0.22  & [1] \\
$r$  &  21.90$\pm$0.14  & [1] \\
$i$  &  22.03$\pm$0.23  & [1] \\
$m_{\mathrm{11200~{\AA}}}$  &  20.20$\pm$0.20  & this work, Sect.~\ref{sect:sinfoni}. \\ 
$m_{\mathrm{13500~{\AA}}}$  &  19.90$\pm$0.25  & this work, Sect.~\ref{sect:sinfoni}. \\ 
$K$  &  19.16$\pm$0.13  & [2] \\ 
\hline
\end{tabular}
\caption{The AB magnitudes of the $z=0.7163$ galaxy, not corrected for
Galactic reddening ($E_{B-V}=0.024$, \citet{schlafly11}).
  References: [1] \citet{rao11}, [2] \citet{hewett07}.}
\label{tab:phot}
\end{table}

\begin{figure}
\begin{center}
\includegraphics[width=0.52\textwidth]{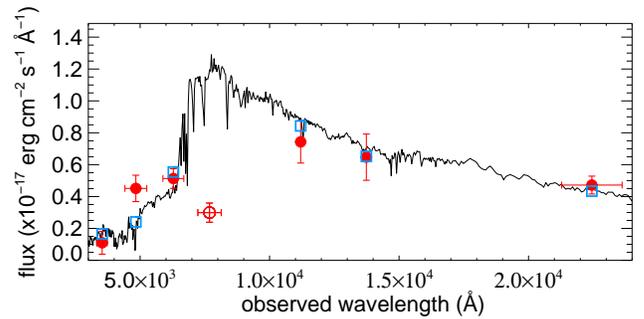}
\end{center}
\vskip -0.5 cm
\caption{The best fit template fit to the galaxy's SED is an
  instantaneous burst with an age of 0.5 Gyr, a metallicity that is
  2.5 times the solar metallicity, and a reddening of $A_V=0.4$ mag.
  The $i$-band shown by the outlined circle was not used in the fit
  because of large QSO subtraction residuals. The blue squares
  represent the expected flux, given the template model and the
  transmission curves of the different filters.}
\label{fig:sedfit}
\end{figure}

We obtain a best-fit stellar mass of 
log$[{\rm M}_*/{\rm  M}_\odot]=10.80^{+0.07}_{-0.14}$, as illustrated in
Fig.~\ref{fig:sedfit}, with a reddening of $A_V=0.4^{+0.3}_{-0.3}$~mag,
2.5 times solar metallicity, and an age of $\approx 0.5$~Gyr,
assuming an instantaneous burst of star formation. To derive the 68\%
confidence intervals for the results, we assume a Gaussian error
distribution, propagate the magnitude errors, and perform SED fits
to $10^3$ random realisations.  Including the uncertain $i$-band data
point in the fit (shown as a red open circle in Fig.~\ref{fig:sedfit})
significantly increases the $\chi^2$ for the best fit, but yields a
very similar derived stellar mass,
log$[{\rm M}_*/{\rm M}_\odot]=10.94^{+0.05}_{-0.08}$.

\begin{table*}
\caption{Physical parameters of the DLA host galaxy g1323. We
provide parameters determined in this work and in addition parameters
from the literature when relevant for the discussion of our new results.}
\begin{tabular}{llr}
\hline
\hline
Item   & value  & reference(s) \\
\hline
log[${\rm M}_*$/M$_\odot$]   & $10.80^{+0.07}_{-0.14}$      & this work$^3$, Sect.~\ref{sect:sedfit} \\
log[M$_{\rm mol-gas}$/M$_\odot$]$\dagger$   & $10.37 \pm 0.04$  & this work, Sect.~\ref{sect:molmass} \\

log[M$_{\rm dyn}$/M$_\odot$]  & $\gtrsim 11.5 $  & this work, Sect.~\ref{sect:W20W50} \\

$\log [{\rm M}_{\mathrm{dust}}/{\rm M}_\odot]$ & $8.45^{+0.10}_{-0.30}$ & this work, Sect.~\ref{sect:mdust} \\

$\log [{\rm L}_{\mathrm{dust}}/{\rm L}_\odot]$ & $11.15\pm 0.20$
& this work, Sect.~\ref{sect:mdust} \\

${\rm T}_{\mathrm{dust}}$ & $21.5\pm 2$ K
& this work, Sect.~\ref{sect:mdust} \\

$z_{abs}$ $z_{\tau:50}$      & $0.71612$ 	    & this work, Sect.~\ref{sect:uves}\\
$z_{abs}$ $z_{\tau:max}$     & $0.71625$            & this work, Sect.~\ref{sect:uves}\\
$z_{em}$ CO(2-1)          & $0.71645 \pm 0.00009$   & this work, Sect.~\ref{sect:W20W50} \\
$z_{em}$ \fion{O}{ii} 	     & $0.7170 \pm 0.0006$  & this work, Sect.~\ref{sect:fors}\\
$z_{em}$ H$\alpha$           & $0.7171 \pm 0.0001$  & this work, Sect.~\ref{sect:sinfoni}\\
PA ($K$-band)                & 24.5$^{\circ}$       & a  \\
PA (dust)                    & $(17\pm8)^{\circ}$   & this work, Sect.~\ref{sect:centroid}\\
PA (CO)                      & $(32\pm14)^{\circ}$  & this work, Sect.~\ref{sect:centroid}\\
PA ($J$-band)                & ($29\pm7)^{\circ}$   & this work, Sect.~\ref{sect:sinfoni}\\
PA (H$\alpha$)               & ($32\pm8)^{\circ}$   & this work, Sect.~\ref{sect:sinfoni}\\
Impact parameter ($K$-band)  & 1\farcs25 = 9.1 kpc     & a  \\
Impact parameter (dust)      & 1\farcs46$\pm$0\farcs2  & this work, Sect.~\ref{sect:centroid}\\
Impact parameter (CO)        & 1\farcs03$\pm$0\farcs3  & this work, Sect.~\ref{sect:centroid}\\
Impact parameter ($J$-band)  & 1\farcs3$\pm$0\farcs1   & this work, Sect.~\ref{sect:sinfoni}\\
Impact parameter (H$\alpha$) & 1\farcs2$\pm$0\farcs1   & this work, Sect.~\ref{sect:sinfoni}\\
CO flux density              & $0.50 \pm 0.05$ Jy \kms & this work, Sect.~\ref{sect:molmass}\\
\fion{O}{ii} flux            & $2.8\pm 0.3\times 10^{-17}$ erg cm$^{-2}$ s$^{-1}$& this work$^1$, Sect.~\ref{sect:fors}\\
H$\alpha$ flux               & $12.0\pm3.9 \times 10^{-17}$ erg cm$^{-2}$ s$^{-1}$ & this work, Sect.~\ref{sect:sinfoni}\\
\fion{N}{ii}$\lambda\lambda$6584 flux & $5.6\pm1.8\times 10^{-17}$ erg cm$^{-2}$ s$^{-1}$ & this work,  Sect.~\ref{sect:sinfoni}\\
W50 (CO line FWHM)           & $647 \pm 54$~\kms\      & this work, Sect.~\ref{sect:W20W50} \\
W20 (CO line)                & $652 \pm 54$~\kms\      & this work, Sect.~\ref{sect:W20W50} \\
inclination angle ($i$)      & 40-48$^o$               & this work, Sect.~\ref{sect:W20W50} \\
\fion{O}{ii} line FWHM       & $< 300$~\kms\           & this work, Sect.~\ref{sect:fors} \\
H$\alpha$ line FWHM          & $238 \pm 34$~\kms\      & this work$^5$, Sect.~\ref{sect:sinfoni} \\
log[$N$(\ion{H}{i})/cm$^{-2}$]   &  $20.21^{+0.21}_{-0.18}$, $20.54^{+0.16}_{-0.15}$    & b, c \\
Absorption metallicity, [Zn/H] & $0.61 \pm 0.20$, $0.4 \pm 0.3$ & d, Sect.~\ref{sect:uves} \\ 
Galaxy abundance, 12+log(O/H) & $9.20 \pm 0.09$ ($\approx 3.2\times$ solar)             & this work, Sect.~\ref{sect:abundance}\\ 
$\Delta v_{90}$              &  $141 \pm 2$~\kms\           & this work, Sect.~\ref{sect:uves} \\
SFR (UV)                     &$0.43\pm 0.23$ M$_\odot$ yr$^{-1}$        & this work, Sect.~\ref{sect:sfr}  \\
SFR (\fion{O}{ii})           & $0.51 \pm 0.05$~M$_\odot$ yr$^{-1}$ & this work$^2$, Sect.~\ref{sect:sfr}  \\
SFR (H$\alpha$, SINFONI)     & $1.6\pm 0.4$ M$_\odot$ yr$^{-1}$
& this work, Sect.~\ref{sect:sfr}  \\

SFR (dust obscured)   & $4.5_{-2.6}^{+4.4}$ M$_\odot$ yr$^{-1}$        & this work, Sect.~\ref{sect:mdust}  \\
SFR (total)   & $6.1_{-2.6}^{+4.4}$ M$_\odot$ yr$^{-1}$        & this
work, Sect.~\ref{sect:fund-plane} \\ 
A$_V$                        & $0.4^{+0.3}_{-0.3}$           & this work$^3$, Sect.~\ref{sect:sedfit}\\
$v_{rel}$(CO-abs)      & $58 \pm 16$ \kms\   & this work  \\
$v_{rel}$(\halpha-abs) & $171 \pm 18$ km s$^{-1}$          & this work, Sect.~\ref{sect:uves}\\
$R_e$   &  4.0 kpc                                         &  a \\
\hline
\end{tabular}
\flushleft
$^\dagger$~Assuming $\alpha_{\rm CO} = 4.2$~M$_\odot$~(K~\kms~pc$^2$)$^{-1}$ and $r_{21} = 0.6$.  \\
$^1$ Corrected for Galactic extinction, but not for intrinsic extinction
in the galaxy. The FORS spectrum was obtained in non-photometric
conditions, and slit losses are estimated to be $\approx 50$\%.
A correction of at least a factor of two upwards is likely needed.\\
$^2$ Based on \fion{O}{ii} flux so notes above (under $^1$) still apply.
Assuming a Chabrier IMF. \\
$^3$ Based on SED fit. \\
$^4$ References: 
[a] \citet{chun10};
[b] \citet{rao06};
[c] \citet{khare04};
[d] \citet{peroux06}
\\

$^5$ Corrected for resolution. \\
\label{tab:summary}
\end{table*}

\subsubsection{Star formation rates}
\label{sect:sfr}

The galaxy has been previously suggested to be devoid of emission lines
and most likely a 'red-and-dead' type. With our detection of multiple
emission lines and 140 GHz continuum, we are now able to estimate the
SFR using several
tracers: the UV continuum emission, the \fion{O}{ii} and H$\alpha$
lines, but also the obscured SFR can be determined
(see Sect.~\ref{sect:mdust}). We convert the measured flux densities and
line fluxes to luminosities in the $\Lambda$CDM model adopted here,
and then to SFRs, using the conversion factors in \citet{kennicutt98}.
This calibration assumes a Salpeter IMF; a simple downward correction
factor of 1.8 is finally used to shift to the assumption of a
Chabrier IMF, for consistency with our SED fitting.

The $u$-band central transmission corresponds to a restframe
wavelength of $\sim$2000{\AA} which implies an unobscured
SFR$_{\mathrm{UV}}=0.43\pm0.23$~M$_{\odot}$~yr$^{-1}$. The large
uncertainty reflects the large $u$-band magnitude error.

The \fion{O}{ii} emission line yields
SFR$_{\mathrm{[\ion{O}{II}]}}=0.51\pm0.05$~M$_{\odot}$~yr$^{-1}$,
while the H$\alpha$ line yields
SFR$_{\mathrm{H\alpha}}=1.6\pm0.4$~M$_{\odot}$~yr$^{-1}$.  Apart from
the fact that H$\alpha$ and \fion{O}{ii} trace SFRs on slightly
different times scales, which could lead to this difference, we also
stress that the H$\alpha$ emission is measured in a larger aperture
than that used for the \fion{O}{ii} emission. In addition, the
\fion{O}{ii} flux might be underestimated due to the non-photometric
observing conditions and will be underestimated due to
intrinsic extinction.

\subsubsection{Dust mass, dust temperature and the obscured SFR}
\label{sect:mdust}

To determine the dust mass in the host galaxy, we use {\sc
magphys} \citep{dacunha08} with the photometry in
Table~\ref{tab:phot}, the ALMA 140~GHz measurement of $53 \pm
17$~$\mu$Jy, and the Herschel photometry in
Table~\ref{tab:herschel_phot}, again fixing the redshift to that of the
galaxy. {\sc magphys} assumes a Chabrier IMF to generate galaxy
templates, which makes the results directly comparable to the output
of {\sc Hyperz}.

The Herschel resolution does not allow us to determine which
fraction of the flux is from the DLA galaxy and which is from
the QSO. In the higher resolution ALMA data we find that the
relative fraction is $0.27/0.73$, and in Sect.~\ref{sect:herschel}
we assumed the same ratio to hold for the Herschel data. Here we
test this assumption. We proceed by forming a series of photometric
tables where we split the Herschel flux $0.10/0.90$, $0.20/0.80$, ...
$0.90/0.10$ in steps of 0.10. For each table we derive the best
fit model templates, and compute the combined $\chi^2$ from the
Herchel and ALMA data. The best fit is found where 0.30 of the
Herschel flux comes from the DLA galaxy, supporting the
assumption used in Sect.~\ref{sect:herschel}. With this assumption 
the maximum likelihood for the dust mass is found to be $\log [{\rm
    M}_{\mathrm{dust}}/{\rm M}_\odot]= 8.45^{+0.10}_{-0.30}$, with a
luminosity of $\log [{\rm L}_{\mathrm{dust}}/{\rm
    L}_\odot]=11.15\pm0.20$.  The maximum likelihood peaks at a stellar
mass of $\log [{\rm M}_*/{\rm M}_\odot] =10.85\pm0.20$, in full
agreement with our result in Sect.~\ref{sect:sedfit} 
($10.80^{+0.07}_{-0.14}$). The fit yields an
obscured SFR of $4.5^{+4.4}_{-2.6}$~M$_{\odot}$~yr$^{-1}$
and a best fit dust temperature of 21.5$\pm$2 K in the ambient
interstellar medium.

In summary, the three optical/UV tracers yield low SFRs, in the range
$\approx 0.4 - 1.6$~M$_\odot$~yr$^{-1}$. While the obscured SFR inferred
from the fit to the dust emission is a little higher, it is still
relatively low for a galaxy with this high stellar mass;
its specific SFR is = SFR/${\rm M}_*
\approx 10^{-11}$~yr$^{-1}$, which places it in the red sequence of
passively evolving galaxies in the correlation between SFRs and
stellar masses \citep{wuyts11}.

\subsubsection{Emission line metallicity}

\label{sect:abundance}
Having detected both H$\alpha$ and \fion{N}{ii} emission lines, we can
use their line flux ratio $N2=\log f$(\fion{N}{ii}~6584)/$f$(\halpha)
to infer the oxygen abundance of the host
\citep[e.g.][]{pettini04,kewley08}.
We note that the relation between $N2$ and the
metallicity is known to saturate at high metallicities
\citep[e.g.][]{steidel14}, and that one has to rely on extrapolations
above solar metallicity \citep{pettini04}.
Here, we use the calibration relation of \citet{maiolino08},
which has been calibrated
to higher-metallicity SDSS galaxies, using photoionization models
\citep{kewley02}. Equally important, this calibration has been used to
tie the conventional galaxy mass-metallicity relation
\citep{tremonti04} to the mass-metallicity relation in DLAs
\citep{moller13,christensen14}.

From the $N2$ index, we infer 12+log(O/H)~$= 9.20 \pm 0.09$.
In addition to the measurement error, the $N2$ calibration itself
has an intrinsic spread of $\approx 0.2$ dex. The solar
oxygen abundance is 12+log(O/H)~=~8.69 \citep{asplund09}, yielding an
oxygen metallicity of [O/H]~$= +0.51 \pm 0.09$ (not including
the calibration uncertainty), i.e. 3.2 times the
solar value. This is in excellent agreement with both the independent
result from the best fit SED model which had 2.5 times the
solar value, and with the DLA absorption metallicity
(Sect.~\ref{sect:uves}).

\section{Discussion}

In the vast array of possible techniques to select high-$z$ galaxies
(e.g. Lyman-break, Lyman-$\alpha$ emission, sub-mm emission, QSOs,
ULIRGs, radio galaxies), the selection via damped Lyman-$\alpha$
absorption has always held a unique position. The use of absorption
towards a background QSO implies that this selection is not subject to
the usual luminosity bias that affects all emission-based methods,
rather it is solely a function of the HI absorption cross-section.
Consequently, some of the galaxies identified via the DLA approach
may well not be identifiable by any other means; however, there does
exist some overlap with traditional flux-limited samples, and this
overlap allows one to use DLA samples to extend galaxy scaling
relations out beyond the boundaries of emission-selected, flux-limited
samples. Nevertheless, one of the critical questions in DLA studies has
long
been: ``Is selection via damped Lyman-$\alpha$ absorption simply an
alternative way to identify the same galaxies that are detected via
flux-limited surveys, or does this selection also allow detection
of galaxies that are missed in other surveys?''.
Several papers have addressed this question over the past decades
considering all previously observable properties, but one
important property, the molecular gas content, has so far not been
possible to study for DLA galaxies. Now, with ALMA in operation, we
are finally able to start filling in this missing section of
parameter space.

\begin{figure*}
\hskip -0.6 cm
\includegraphics[width=9.0 cm]{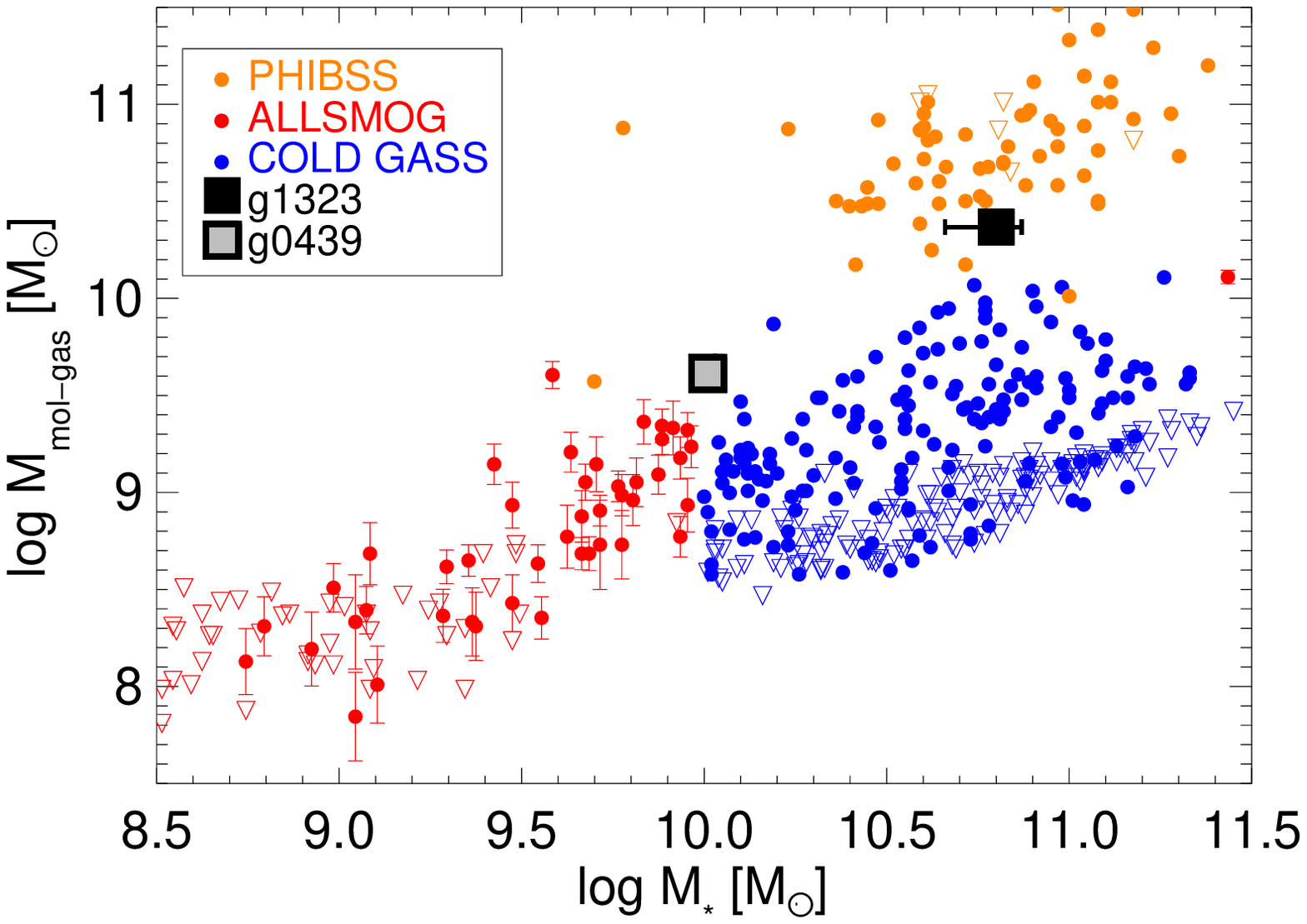}
\includegraphics[width=9.0 cm]{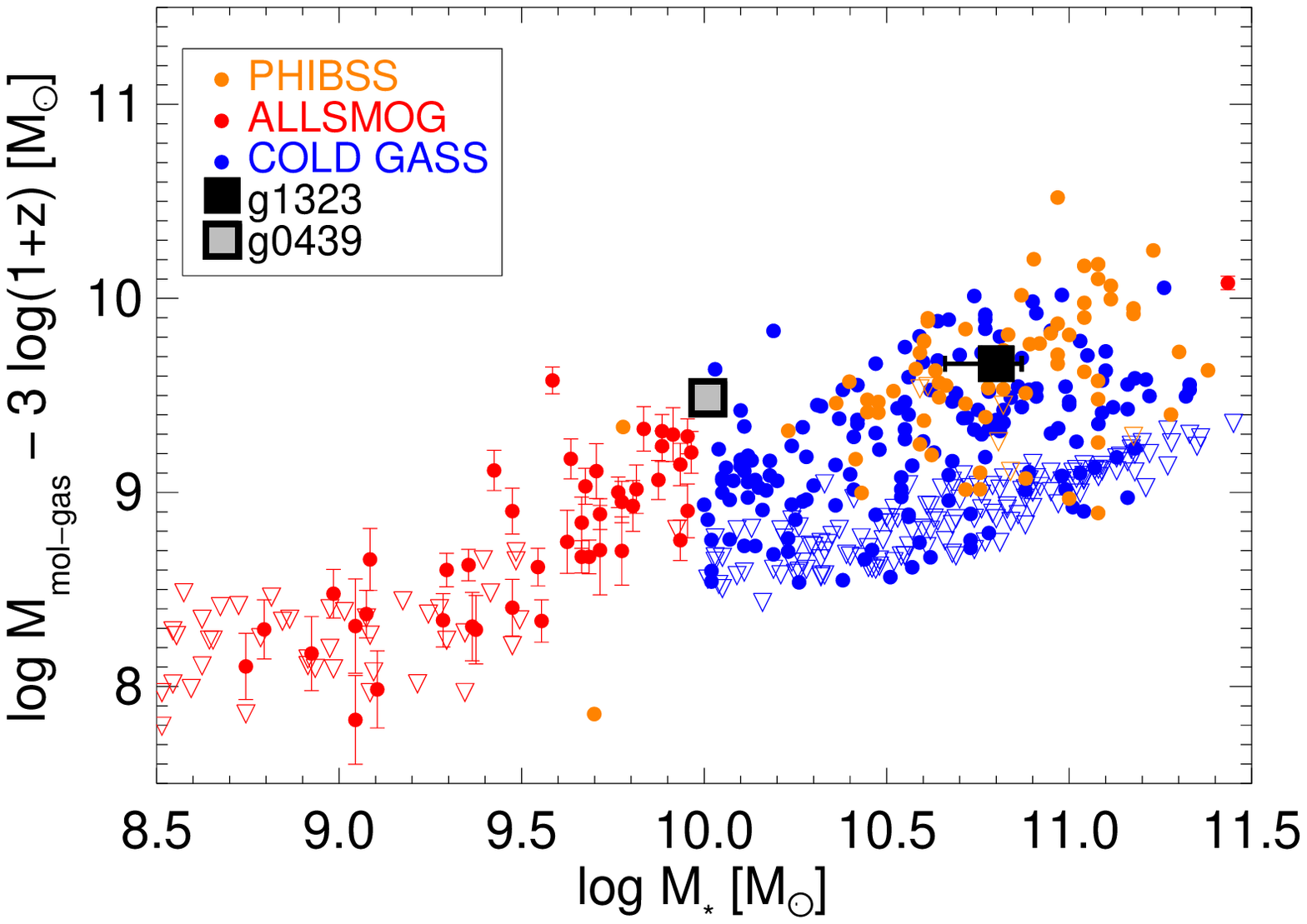}
\caption{The molecular gas mass plotted against the stellar mass for
the first two DLA galaxies detected in CO emission
(redshifts 0.101 and 0.7163), compared to three
literature samples: ALLSMOG, COLD GASS, and PHIBSS (at redshifts
$\approx 0.02$, $\approx 0.04$, and $\approx 1.2 \:,\: 2.2$, respectively).
The same CO-to-H$_2$ conversion factor, $\alpha_{\rm CO} = 4.2 \: 
{\rm M}_\odot$~(K~\kms~pc$^2$)$^{-1}$, was used for all galaxies in
both panels.  Upper limits on M(H$_2$) are indicated by open triangles
pointing down.  In the left panel, no correction has been applied for
the redshift evolution of the relation between molecular gas mass and
stellar mass; in the right panel, a redshift correction of $(1+z)^{-3}$ 
has been applied to the molecular gas mass. See text for discussion.
}
\label{fig:H2mass30}
\end{figure*}

\begin{table}
\caption{Properties of g0439}
\begin{tabular}{llr}
\hline
\hline
Item   & value  & reference \\
\hline
log$[{\rm M}_*/{\rm M}_\odot]$     & $10.01 \pm 0.02$  & a \\
log$[{\rm M}_{\rm mol-gas}/{\rm M}_\odot]$ & $9.60 \pm 0.1$    & b \\
W50 (CO line FWHM)		   & $283\pm21$~\kms\  & c \\
W20 (CO line)			   & $285\pm21$ \kms\  & c \\
inclination angle ($i$)            & $67^{+6}_{-5}$ deg       & b \\
\hline
\end{tabular}
\flushleft
[a]~\citealp{christensen14}; [b]~\citealp{neeleman16};
[c]~this work.
\\
 \label{tab:0439}
 \end{table}

Above we have reported the discovery of a CO line emitting galaxy
(ALMA J132323.81-002154.1) at $z_{CO} = 0.71645$ and at the position
of a known optical candidate DLA host.  
In this section we now compare the properties of this DLA galaxy
to the properties of emission-selected galaxy samples. In particular,
we ask if it lies within the range of the scaling relations known from
flux-selected samples. Where possible, we also include our previous
detection of the absorbing galaxy towards PKS~0439-433 =
QSO~J0441-4313 \citep{neeleman16}. 
The \hi\ column density of the absorber in PKS~0439-433
is below the classic definition of a DLA, but for simplicity we
shall in what follows refer to both of the galaxies simply as ``DLA
galaxies'' but keeping in mind the sub-DLA nature of the absorber
towards PKS~0439.

\subsection{The relation between molecular gas mass and stellar mass}

Fig.~\ref{fig:H2mass30} plots the molecular gas mass of the two
absorbing galaxies versus their stellar mass, along with data from
three comparison galaxy samples: COLD GASS, at $z \approx 0.02$
\citep[][in blue]{saintonge11}, ALLSMOG at $z \approx 0.04$
\citep[][in red]{bothwell14,cicone2017}, and PHIBSS at $z \approx 1.2$ and
$\approx 2.2$ \citep[][in yellow]{tacconi13}. For convenience we
use the same CO-to-H$_2$ conversion factor, $\alpha_{\rm CO} =
4.2$~M$_\odot$~(K~\kms~pc$^{2}$)$^{-1}$ for all galaxies in the
figure. The left panel shows that the COLD GASS and ALLSMOG samples
lie on the well-known
correlation between molecular gas mass and stellar mass. Conversely,
the two absorbers lie at or above the upper boundary of this wide
relation. For easy reference in this figure, and in the remainder of
the paper, we use the short-hand notation g0439 and g1323
for the two galaxies.
In Table~\ref{tab:summary}, we list the known properties of g1323;
Table~\ref{tab:0439} provides a summary of the relevant properties of
g0439.

Galaxies from the PHIBSS sample are seen to have systematically
larger gas masses than the two absorbers, as well as than the flux
selected samples in the nearby Universe.  
However, to obtain a fair comparison between the galaxies at different
redshifts, we must take the redshift evolution of this scaling
relation into account. The simplest view is that the relation between
molecular gas mass and stellar mass scales (empirically) as
$(1+z)^{-3}$ between redshifts 0 and 2.5 \citep{genzel15}; this
empirical correction is applied in the right panel of
Fig.~\ref{fig:H2mass30}. The figure shows that the high-$z$ PHIBSS
points have now moved down to perfectly overlap with the two low-$z$
galaxy samples.  The two absorbers remain in the upper part of the
distribution, but now lie within the spread of the distribution of the
emission-selected galaxies. We note that both absorbers appear to be
rich in molecular gas; notably, g0439 lies on the upper boundary
of the distribution.

\subsection{The SFR, molecular gas fraction, and stellar mass relations}
\label{sect:fund-plane}

\begin{figure}
\hskip -0.6 cm
\includegraphics[width=9.0 cm]{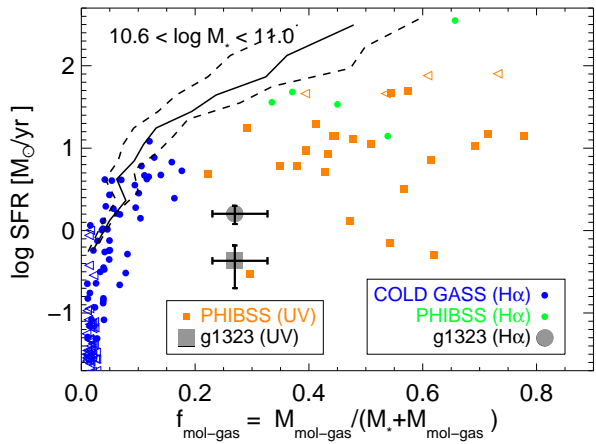}
\caption{SFR vs molecular gas fraction
  $f_{\rm mol-gas} = {\rm M}_{\rm mol-gas}/({\rm M}_{\rm mol-gas} + {\rm
  M}_*)$. Following the
  fundamental relation between M$_*$, $f_{\rm gas}$ and SFR
  we plot only galaxies in a narrow stellar mass range centred at
  log$[{\rm M}_*/{\rm M}_\odot]=10.80$ (full curve), with the mass range $10^{10.6}
  \leq {\rm M}_*/{\rm M}_\odot \leq 10^{11.0}$ (dashed curves).
  We plot UV- and \halpha
  -based SFRs with different symbols. Orange squares mark $z\approx 1.2$
  UV-based SFR, green circles mark $z\approx 2.2$ \halpha-based SFR 
  determinations. Triangles
  pointing to the left are upper detection limits on $f_{\rm mol-gas}$.}
\label{fig:DLA1323Ha}
\end{figure}

\begin{figure}
\hskip -0.6 cm
\includegraphics[width=9.0 cm]{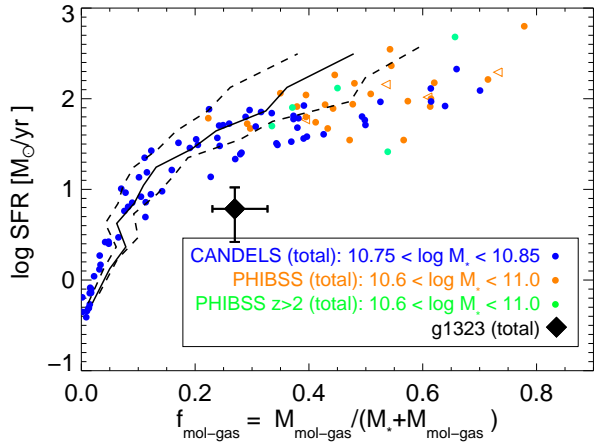}
\caption{Same as Fig.~\ref{fig:DLA1323Ha}
  but here we plot total SFR.  
  Again the orange and green dots are objects from
  the PHIBSS survey while the black lines mark the fundamental
  relation shown in Fig.~\ref{fig:DLA1323Ha}. For COLD GASS we do not
  have total SFRs and instead we plot in blue galaxies from the
  CANDELS sample in a very narrow stellar mass range
  $10^{10.75} \leq {\rm M}_*/{\rm M}_\odot \leq 10^{10.85}$.  
  The flux limited samples are seen to follow a relation similar to
  that previously found for total gas mass (black curve), but shifted
  increasingly to the right for larger $f_{\rm mol-gas}$.
  g1323 is again
  seen to have a very low SFR for a galaxy with its stellar mass and
  large molecular gas mass fraction.} 
\label{fig:DLA1323IR}
\end{figure}

The other view on the influence of redshift on scaling relations is
that of ``fundamental relation invariance''. In this view, the values
of physical parameters describing the galaxies are tied to
each other to only allow galaxies to fill small sub-sets of physical
parameter space, and galaxy samples then ``drift'' through this
parameter space, yielding an apparent redshift evolution in individual
parameters.
\citet{santini14} formulated such a fundamental relation
between the SFR, the gas fraction ($f_{\rm gas}= M_{\rm
gas}/(M_* + M_{\rm gas})$, where $M_{\rm gas}$ is the total
gas mass), and the stellar mass, with the claim that all galaxies,
at all redshifts (out to $z \approx 2.5$, the upper limit of the
redshifts of the galaxies of their sample), must lie on this relation.  
In order to test if a similar relation exists for molecular gas
mass alone ($f_{\rm mol-gas}= M_{\rm mol-gas}/(M_* + M_{\rm mol-gas})$),
and in particular to test if g1323 then lies on this relation, we
first use the
same comparison samples as above, but now cut out a thin slice in
stellar mass of each sample, centred on the $M_*$ of g1323.  
In figures~\ref{fig:DLA1323Ha} and \ref{fig:DLA1323IR} we plot
SFR vs $f_{\rm mol-gas}$ of galaxies within the mass range
$10^{10.6} \leq {\rm M}_*/{\rm M}_\odot \leq 10^{11.0}$. Several
different indicators for the SFR are available, applicable at
various wavelengths, and in order to provide appropriate and
unbiased comparisons
we use different symbols for UV- and H$\alpha$-based SFR in
Fig.~\ref{fig:DLA1323Ha}, while total SFRs corrected for the dust
obscured component are plotted in Fig.~\ref{fig:DLA1323IR}.
For comparison we also plot the relevant total gas mass
relations within which the galaxies should lie
\citep[computed from equation~9 and Table~1 of][]{santini14}.
For completeness we note here that the calibration of the
relation as parameterised by \citet{santini14} is based on a Salpeter
IMF. For an internally consistent comparison, we have converted their SFR
values to a Chabrier IMF, by subtracting 0.24~dex from their reported
log$[{\rm M}_*/{\rm M}_\odot]$ and 0.15~dex from their log[SFR].

In order to obtain SFRs for the three comparison samples we proceeded as
follows. COLD GASS galaxies were cross-matched with the SDSS database,
and H$\alpha$-based SFRs were extracted from aperture corrected SFRs
\citep{brinchmann04}\footnote{SFRs are obtained from the JHU-MPA
value-added catalog
https://www.sdss3.org/dr9/algorithms/galaxy\_mpa\_jhu.php}. For ALLSMOG
galaxies \citet{cicone2017}
(ALLSMOG website\footnote{http://www.mrao.cam.ac.uk/ALLSMOG/}) 
list the SFRs (again H$\alpha$ and based on SDSS database values), and 
we also obtained H$\alpha$-based SFRs for part of the PHIBSS sample from
\citet{erb06}. For the remaining galaxies in the COSMOS and
GOODS-N fields we obtained estimates of the UV SFRs from the Rainbow
database\footnote{https://rainbowx.fis.ucm.es/Rainbow\_Database/Home.html},
which reports the UV-based SFRs derived from SED fits with FAST (Kriek
et al. 2009). \citet{tacconi13} report a total SFR from either
extinction corrected H$\alpha$ emission (filled green circles in
Fig.~\ref{fig:DLA1323IR}), or from the sum of UV- and IR-based SFRs
(filled yellow circles in Fig.~\ref{fig:DLA1323IR}). The systematic
uncertainties are reported to be $\pm0.15$ dex.  We compute the
total SFR of g1323 by adding our H$\alpha$-based SFR and the dust
obscured SFR and find
SFR$_{\rm total}=6.1^{+4.4}_{-2.6}$~M$_\odot$~yr$^{-1}$.

From Fig.~\ref{fig:DLA1323Ha} we first note that the UV-based SFRs (square
symbols) all fall far below the relation, and that g1323 is in the
very low range of this wide distribution. Considering now only the
H$\alpha$-based SFRs (filled circles), they are all seen to lie closer
to the relation, and are likely much better SFR indicators for galaxies
with large molecular gas mass fractions.
It is seen that most points fall well below the corresponding
relation for total gas mass, and increasingly so for larger gas mass
fractions. Still, even for $f_{\rm mol-gas}$ as large
as 0.4-0.5 the PHIBSS galaxies fall only marginally below the relation.
In contrast, the H$\alpha$-based SFR of g1323 falls significantly
below it (note that the flux-based uncertainty is linear in SFR, but
we are plotting log(SFR) in the figures).
While g1323 obviously has a large $f_{\rm mol-gas}$ compared to the
COLD GASS galaxies, it is still surprising that it falls so far below
the relation followed by the low $f_{\rm mol-gas}$ galaxies, and even
far below the PHIBSS galaxies that have larger $f_{\rm mol-gas}$.  
Either g1323 could have an even larger hidden SFR than the PHIBSS
galaxies, or it simply has an extremely low SFR for its stellar mass
and gas fraction.

To find out which is true, we therefore plot, in
Fig.~\ref{fig:DLA1323IR}, the ``total SFR'' corrected for dust.
We do not have the information to compute the dust correction for the
COLD GASS galaxies but total SFRs are available for both sub-samples
of the PHIBSS galaxies. The corrected PHIBSS sample is seen now to lie
closer to the \citet{santini14} relation but it still lies somewhat
below and with a scatter slightly larger than the mass slice would
predict. In order to obtain a better evaluation of how the relation
based on $f_{\rm mol-gas}$ compares to that of the total gas mass
relation, we also include CANDELS galaxies from
\citet{popping15} for which total SFRs are also known. The
molecular gas mass for the CANDELS galaxies has not been measured,
rather it has been inferred from a suite of optical data and the
model presented in \citet{popping15}. It is seen that for 
$f_{\rm mol-gas} < 0.2$ the CANDELS sample follows the \citet{santini14}
relation well, whereas it for larger $f_{\rm mol-gas}$ connects
perfectly with the PHIBSS sample. The CANDELS sample is very large
allowing us to select an even narrower mass slice
($10^{10.75} \leq {\rm M}_*/{\rm M}_\odot \leq 10^{10.85}$) than for
PHIBSS. While care must be exercised in not over interpreting the
partly model based CANDELS data points, the general trend of the
relation from small to large $f_{\rm mol-gas}$ appears now
to be well established. The scatter of the CANDELS sample is seen to
be smaller than for the PHIBSS sample, which was expected because the
CANDELS galaxies are picked from a narrower mass range. g1323 is
again seen to fall well below both samples.
We conclude this discussion summarising our two main results:
({\it i}) there is indeed a fundamental relation for
($f_{\rm mol-gas}$, SFR, $M_*$) similar to that of total gas mass
fraction but shifted a little towards larger $f_{\rm mol-gas}$ and,
({\it ii}) g1323 lies well below this relation.  
It therefore appears that g1323, seen in this
observed parameter space and stellar mass range, does not have
counterparts in flux-selected samples. It is noteworthy that
\citet{neeleman16} reported a low SFR and large gas consumption
timescale for g0439.

There are ways to apply
reasonable assumptions to force g1323 onto the relation of course,
but before we turn to those we emphasize that such assumptions have
also not been applied to any of the objects of the comparison samples,
so at face value, with the zero'th order assumptions, g1323 is
significantly inconsistent, both considering H$\alpha$-based and total
SFR, with the relation of the flux-selected samples.

For g1323 we found (Sect.~\ref{sect:mdust}) that there is indeed
evidence for some obscured SFR (as included in Fig.~\ref{fig:DLA1323IR}).
One could hypothesize that even more SFR is hidden by dust, thereby
pushing the point further upwards towards
the relation. This hypothesis is not supported by the data analyzed
in Sect.~\ref{sect:mdust}. Alternatively one could assume a lower
value for $\alpha_{\rm CO}$ for g1323, which would push the point
towards the left. Assuming e.g. the extreme example
(Sect.~\ref{sect:molmass}) of a starburst galaxy, with
$\alpha_{\rm CO} \approx 1$ and $r_{21} \approx 1$, would imply
$f_{\rm mol-gas} = 0.05$. The assumption of such a
highly obscured starburst is also not supported by our observations, as
described in Sect.~\ref{sec:gastodust}.


\begin{figure}
\hskip -6 mm
\includegraphics[width=0.52\textwidth]{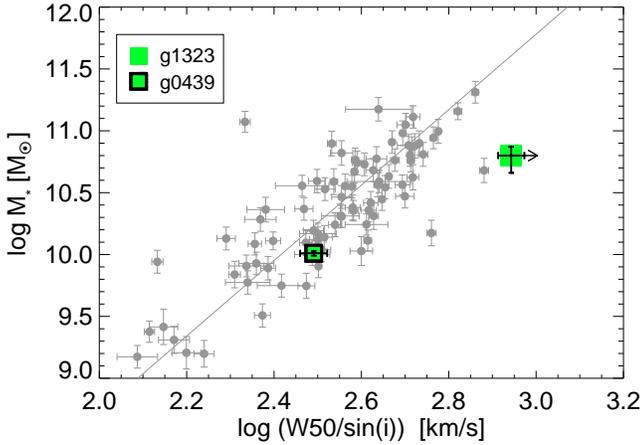}
\caption{Comparison to the CO-Tully-Fisher relation of the COLD GASS
sample from \citet{tiley2016} (small grey dots). For g1323 we have
used the conservative estimate $i=48^o$ (Sect.~\ref{sect:W20W50}). The
inclination could be smaller, and the arrow marks where it would move
for an inclination of e.g. $i=40^o$.}
\label{fig:MvsW20}
\end{figure}

\subsection{CO-Tully-Fisher relation}
\label{sec:TF}

In Fig.~\ref{fig:MvsW20} we compare our absorbing
galaxies to the COLD GASS Tully-Fisher relation
\citep[figure 2b in][]{tiley2016}. For g0439 we fit the same
double-horned profile as for g1323 (Sect.~\ref{sect:W20W50}) and
find the values for W50 and W20 listed in Table~\ref{tab:0439}.
Both galaxies are seen to fall to the right of
the relation (i.e. they have large velocity widths for their
stellar masses). g0439 is fully consistent with the scatter of the
COLD GASS sample, but g1323 is seen to fall significantly right of
the relation.

As discussed in Sect.~\ref{sect:kinematics} there is some evidence
that part of the CO line profile (the lowest redshift peak) may be
due to a gas component which is separate from the rest of the galaxy.
If we instead use W50 = 378 \kms\ from the ``two object fit''
(Sect.~\ref{sect:W20W50}) then the green square moves left by 0.23 dex
in Fig.~\ref{fig:MvsW20}, which brings it
in agreement with the COLD GASS distribution.  The ``outlier'' nature
of g1323 in the TF relation would therefore find a simple explanation
if the existence of this separate component is confirmed.

\subsection{Gas-to-dust mass vs metallicity relation}
\label{sec:gastodust}

\begin{figure}
\begin{center}
\includegraphics[width=0.52\textwidth]{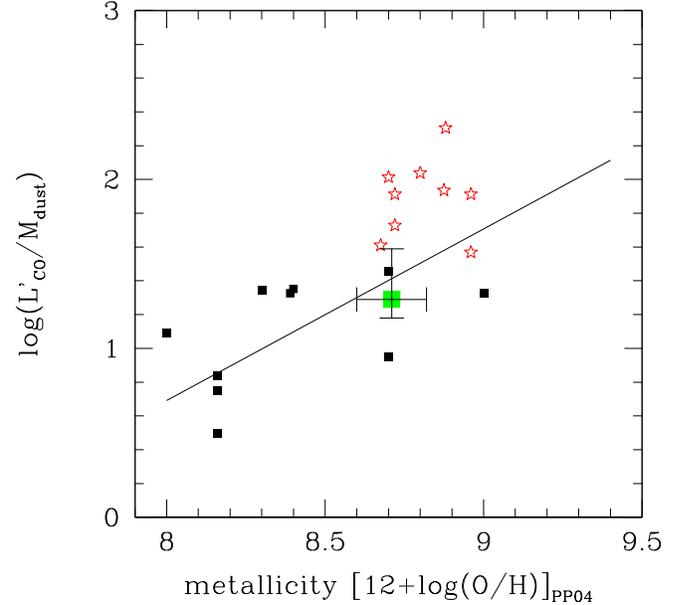}
\end{center}
\vskip -0.5 cm
\caption{Gas-to-dust vs metallicity relation for local group galaxies
(small black squares) and local ULIRGs (red stars). g1323 (large
green square) is seen to lie well within the parameter space of local
group galaxies of similar metallicity (adapted from \citet{magdis11}).
}
\label{fig:gastodust}
\end{figure}

For local galaxies it is known that the ratio of gas to dust mass is
related to the emission line metallicity of the galaxy. We did not
detect any dust emission continuum in our ALMA data of g0439, but
for g1323 we obtained a detection and inferred a dust mass of
$\log [{\rm M}_{\mathrm{dust}}/{\rm M}_\odot]= 8.45^{+0.10}_{-0.30}$
(Sect.~\ref{sect:mdust}). In Fig.~\ref{fig:gastodust} we now
compare g1323 (large green square) to the sample of local group
galaxies (small black squares) and local ULIRGs (red stars) compiled
in \citet{magdis11}. In order to compare to the \citet{magdis11}
sample we need to first convert our measured emission line metallicity
to the calibration by \citet{pettini04} and we find
${\rm[12+log(O/H)]}_{\rm{PP04}}=8.71\pm0.11$. As \citet{magdis11}
(their Figure 3, right) we plot $L'_{CO}$ rather than $M_{gas}$ in
order to avoid the additional uncertainty arising from $\alpha_{CO}$.

From Fig.~\ref{fig:gastodust} we first conclude that g1323 is lying
perfectly on this scaling relation for normal flux-selected galaxies.
While it is
marginally consistent with the position of ULIRGs, the fit to the
relation of local galaxies (black line) is much better, a conclusion
which is also supported by our result in Sect.~\ref{sect:mdust} that
this galaxy has a relatively low amount of obscured SFR for its
stellar mass, as well as only modest reddening (A$_V=0.4$, 
Sect.~\ref{sect:sedfit}).

\subsection{Relation to emission-selected galaxy samples}
DLA galaxies make up a sample of galaxies that are almost exclusively
first identified in absorption (i.e. absorption-selected). This is an
important characteristic to keep in mind, because it means that we do
not a priori know if ``absorption selection'' is a differently weighted
process to find the same galaxies as ``flux-limited selection'', or if
it is a way of finding entirely different objects. This question is
not easily answered, and yet it is of fundamental importance in order
to be able to correctly interpret flux-limited galaxy samples, and in
particular to be able to connect them to the vast
reservoir of information gathered via absorption line studies of DLAs.
Finding the answer requires that we identify a sample of DLA galaxies in
emission, and then that we carry out a detailed study to test if those
objects lie within the same parameter space as emission selected galaxies.

\citet{moller02} and \citet{weatherley05} previously addressed this
exact question using a small set of $z=2$-$3$ DLA galaxies observed
in optical and NIR
emission. The conclusion at that time was that all the measured emission
properties were consistent with the conjecture that DLA galaxies are
Lyman Break Galaxies. One of the key purposes of our ALMA survey is now
to extend this fundamental test into the ALMA frequency range,
where we are able to ask the same question for properties related to
the molecular gas. In this paper we have outlined this procedure, and
compared our first two detections to galaxy
gas scaling relations determined from flux limited samples. Here we
provide a short summary of this comparison.
\begin{itemize}
\item
{\bf M$_{gas}$ vs M$_*$:} both galaxies are rich in molecular gas;
notably, g0439 lies on the upper boundary of the known distribution.
Both are within the parameter space of flux limited samples.

\item
{\bf SFR vs molecular gas fraction:} \citet{neeleman16} reported a low
SFR and large gas consumption timescale for g0439. Similarly we have here
found that the SFR of g1323 is significantly below the flux limited
relation for galaxies of similar stellar mass and molecular
gas-mass fraction, both considering purely H$\alpha$-based SFRs, and
total (dust corrected) SFRs.

\item
{\bf Tully-Fisher relation:} the two galaxies have a large CO line
velocity width for their stellar mass. g0439 is here within the standard
relation of the COLD GASS sample, but g1323 is also in this respect
at best marginally within the flux limited relation.

\item
{\bf Gas-to-dust vs metallicity:} we were only able to obtain a dust
mass of g1323. This galaxy lies directly on the known relation of
normal (i.e. non starburst) main sequence galaxies.

\end{itemize}

The answer is therefore not as clear as it was in the earlier works
where only stellar and ionized gas components were considered.
g1323 appears to be the first case of a DLA-selected galaxy that
does not have counterparts in flux-limited samples as it fails our
consistency test in two out of four relations.  This galaxy may simply
be a very unique and rare case, but it may in fact 
represent an ``in between burst'', or ``post starburst'' low SFR phase
of galactic evolution. If such a phase follows an event of strong
ejection of gas (both molecular and neutral), then the ejected
molecular kinematic component could explain the large CO emission line
width, and also explain why galaxies in such a phase are more easily
found as DLA galaxies due to their larger cross section for HI absorption.

Some support for this interpretation may be found in our detailed
analysis of the ALMA data presented here which suggests that {\it (i)}
g1323 consists of two kinematically independent and interacting
sub-clouds and {\it (ii)} the SED fit to g1323 favours an instantaneous
burst of starformation with an age of 0.5 Gyr. The current data do
not rule out the possibility that both clouds may have a stellar
component, but only one of them is visible on the current (ground
based AO) optical image. An HST image should be able to determine
if the additional cloud belongs to a separate and interacting galaxy.

A key conclusion from
our previous study was that ``the strongest molecular
absorption component of g0439 cannot arise from the molecular disk''
\citep{neeleman16} suggesting that also that system has more than one
molecular gas component.
A larger sample must be collected in order to address how common this
type of objects are in DLA selected samples. Our Cycle 3 data are
currently being analyzed and
they will already significantly enlarge our current sample of
detections (Kanekar et al. in prep).

\section{Summary and conclusions}
We have initiated an ALMA survey for CO emission from galaxies known to
harbour DLA or sub-DLA absorbers. In this paper we present an ALMA map of
the field around the known DLA absorber towards the quasar J1323$-$0021.
Based on the detection of CO(2-1) line emission at a redshift of
$z_{\rm CO}=0.71645$, close to that of the absorber
($z_{\rm abs}=0.71625$), we identify
the galaxy responsible for the absorption. This
is the second DLA/sub-DLA galaxy in which CO emission has been detected,
and it is the highest redshift DLA CO detection to date. This
is also the first case of a DLA galaxy identified via molecular gas
emission, and as such it highlights a new and promising route to the
identification of DLA galaxies.
With ALMA we also detect the dust continuum and therefore measure both
the molecular and the dust mass of the host galaxy.

In addition we present new VLT-FORS2 data and we re-analyze archival VLT
SINFONI data cubes, Herschel maps, and VLT UVES and HST STIS spectra.
This large suite of data allows us to dissect the DLA host galaxy in
terms of stellar mass, gas mass and dust mass; and in terms of
kinematical components of the cold (CO) gas, the ionized (H$\alpha$ and
\fion{O}{ii}) gas and the absorbing gas. Also, we are able to determine
the SFR using several different indicators (UV, \fion{O}{ii},
H$\alpha$ and IR).

Our results can be summarized as follows:
\begin{itemize}

\item
The DLA galaxy g1323 is the second most massive DLA-selected galaxy
known with log[${\rm M}_*$/M$_\odot ]= 10.80^{+0.07}_{-0.14}$,
log[M$_{\rm mol-gas}$/M$_\odot ]= 10.37 \pm 0.04$ and
log[M$_{\rm dust}$/M$_\odot ]= 8.45^{+0.10}_{-0.30}$. This also implies
that it is very gas rich, in the upper part of comparable galaxies from
flux limited samples.

\item
The emission line metallicity of g1323 is super solar
($\approx 3.2\times{\rm solar}$), fully consistent with previously
published absorption metallicities. Its gas-to-dust ratio for this
metallicity is perfectly on the relation for normal flux selected
galaxies. It is somewhat below, but within the errors also marginally
consistent with local ULIRGs.

\item
The velocity width of the CO emission is extremely wide for its
stellar mass. It falls outside the scatter of the CO-Tully-Fisher
relation of COLD GASS galaxies. There is some evidence suggesting
that the wide CO line could be due to an additional and separate
(both dynamically and spatially), interacting CO cloud, e.g. a
previous outflow or/and current infall. This cloud could also be
related to a separately, yet undiscovered, interacting galaxy.

\item
We determine the SFR in several different ways and find that for
a galaxy with its stellar mass and molecular gas-mass fraction, its 
SFR is extremely low. With respect to its poor ability to convert
molecular gas into stars this galaxy has no counterparts in
flux limited samples. It is intriguing that a low SFR and large gas
consumption timescale was also found for our first CO-detected
absorber.

\end{itemize}

As in previous (but optically based) studies we perform the fundamental
test if the properties of g1323 are
consistent with the conjecture that DLA-selected galaxies and
flux-selected galaxies are the same, only
selected in different ways. We find that for the first time we have
identified a DLA galaxy that does not appear to have counterparts in flux
limited samples. The galaxy g1323 has too low SFR for a galaxy
with its stellar mass and gas-mass fraction, and it also has
too wide a CO emission line to fit onto the well known
Tully-Fisher relation.

Our survey is ongoing, and
our Cycle 3 data have already provided additional detections.
If more DLA galaxies of this survey in similar ways do not conform to
the relations of emission selected samples, this could have significant
impact on our understanding of the connection between the gaseous and
the stellar component of galaxies, and how they relate to starformation.
Notably the non-disk molecular gas component of DLA galaxies could
form a large previously unexplored reservoir.

\section*{Acknowledgements}
We are grateful to Mark R. Chun for providing the AO $K$-band image,
to Gergely Popping for providing the unpublished CANDELS molecular
masses, and to Giorgos Magdis and Celine Peroux for useful
discussions. JPUF acknowledges
support form the ERC-StG grant EGGS-278202. LC and HR are supported by
YDUN-DFF grant ID 4090-00079. NK acknowledges support from the
Department of Science and Technology via a Swarnajayanti Fellowship
(DST/SJF/PSA-01/2012-13). This paper makes use of the following ALMA
data: ADS/JAO.ALMA\#2013.1.01178.S. ALMA is a partnership of ESO
(representing its member states), NSF (USA) and NINS (Japan), together
with NRC (Canada), NSC and ASIAA (Taiwan), and KASI (Republic of
Korea), in cooperation with the Republic of Chile. The Joint ALMA
Observatory is operated by ESO, AUI/NRAO and NAOJ.
This work has made use of the Rainbow Cosmological Surveys Database, 
which is operated by the Universidad Complutense de Madrid (UCM), 
partnered with the University of California Observatories at Santa 
Cruz (UCO/Lick,UCSC).
\def\aj{AJ}
\def\araa{ARA\&A} \def\apj{ApJ} \def\apjl{ApJ} \def\apjs{ApJS}
\def\apss{Ap\&SS} \def\aap{A\&A} \def\aapr{A\&A~Rev.}
\def\aaps{A\&AS} \def\mnras{MNRAS} \def\nat{Nature} \def\pasp{PASP}
\def\aplett{Astrophys.~Lett.}

\bibliographystyle{mn}
\bibliography{LISEdla2}

\end{document}